%% file: draft.tex
\documentclass[twocolumn]{aastex631}

\usepackage{natbib}
\usepackage{ulem}
\hypersetup{linkcolor=red,citecolor=blue,filecolor=cyan,urlcolor=magenta}
\usepackage{amsmath}
\usepackage{bbding}
\usepackage{graphicx}

\newcommand{\MO}{$M_\odot$}
\newcommand{\ZO}{$Z_\odot$}
\newcommand{\Chandra}{\textit{Chandra}}

\received{}
\revised
\accepted{}
\submitjournal{ApJ}

\shorttitle{Systematic Gas Perturbations in Cool Cores of HIFLUGCS Clusters}
\shortauthors{Ueda et~al.}

\begin{document}
%\Received{}%{yyyy/mm/dd}
%\Accepted{}%{yyyy/mm/dd}
%\Published{}%{yyyy/mm/dd}

\title{Systematic Perturbations of the Thermodynamic Properties in Cool Cores of HIFLUGCS Galaxy Clusters}

%%% begin:list of authors
% Do NOT capitalize all letters in "textsc".

\correspondingauthor{Shutaro Ueda}
\email{sueda@asiaa.sinica.edu.tw}
\author[0000-0001-6252-7922]{Shutaro Ueda}
\affiliation{Academia Sinica Institute of Astronomy and Astrophysics (ASIAA), No. 1, Section 4, Roosevelt Road, Taipei 10617, Taiwan}

\author[0000-0002-7196-4822]{Keiichi Umetsu}
\affiliation{Academia Sinica Institute of Astronomy and Astrophysics (ASIAA), No. 1, Section 4, Roosevelt Road, Taipei 10617, Taiwan}

\author[0000-0003-4195-6300]{FanLam Ng}
\affiliation{Yuanpei College, Peking University, Yi He Yuan Road 5, Hai Dian District, Beijing 100871, China}
\affiliation{Academia Sinica Institute of Astronomy and Astrophysics (ASIAA), No. 1, Section 4, Roosevelt Road, Taipei 10617, Taiwan}

\author{Yuto Ichinohe}
\affiliation{Department of Physics, Rikkyo University, 3-34-1 Nishi-Ikebukuro, Toshima-ku, Tokyo 171-8501, Japan}

\author{Tetsu Kitayama}
\affiliation{Department of Physics, Toho University, 2-2-1 Miyama, Funabashi, Chiba 274-8510, Japan}

\author{Sandor M. Molnar}
\affiliation{Academia Sinica Institute of Astronomy and Astrophysics (ASIAA), No. 1, Section 4, Roosevelt Road, Taipei 10617, Taiwan}

%%% Please use the following style in case that sorting by 
%%% affiliation is impossible. 
%
% \author{%
%   D-Firstname \textsc{D-Familyname}\altaffilmark{1}
%   E-Firstname \textsc{E-Familyname}\altaffilmark{1,2}
%   and
%   F-Firstname \textsc{F-Familyname}\altaffilmark{2}}
% \altaffiltext{1}{Address of Institute}
% \email{ddddd@xxx.xxx.xx.xx}
% \email{eeeee@xxx.xxx.xx.xx}
% \altaffiltext{2}{Address of Institute}

%% `\KeyWords{}' always has to be placed before `\maketitle'.

%\maketitle

\begin{abstract}
We present an ensemble X-ray analysis of systematic perturbations in the central hot gas properties for a sample of 28 nearby strong cool-core systems selected from the HIghest X-ray FLUx Galaxy Cluster Sample (HIFLUGCS). We analyze their cool-core features observed with the {\it Chandra X-ray Observatory}. All individual systems in our sample exhibit at least a pair of positive and negative excess perturbations in the X-ray residual image after subtracting the global brightness profile. We extract and analyze X-ray spectra of the intracluster medium (ICM) in the detected perturbed regions. To investigate possible origins of the gas perturbations, we characterize thermodynamic properties of the ICM in the perturbed regions and characterize their correlations between positive and negative excess regions. The best-fit relations for temperature and entropy show a clear offset from the one-to-one relation, $T_\mathrm{neg}/T_\mathrm{pos}=1.20^{+0.04}_{-0.03}$ and $K_\mathrm{neg}/K_\mathrm{pos}=1.43\pm 0.07$, whereas the best-fit relation for pressure is found to be remarkably consistent with the one-to-one relation $P_\mathrm{neg}=P_\mathrm{pos}$, indicating that the ICM in the perturbed regions is in pressure equilibrium. These observed features in the HIFLUGCS sample are in agreement with the hypothesis that the gas perturbations in cool cores are generated by gas sloshing. We also analyze synthetic observations of perturbed cluster cores created from binary merger simulations, finding that the observed temperature ratio agrees with the simulations, $T_\mathrm{neg}/T_\mathrm{pos}\sim 1.3$. We conclude that gas sloshing induced by infalling substructures plays a major role in producing the characteristic gas perturbations in cool cores. The ubiquitous presence of gas perturbations in cool cores may suggest a significant contribution of gas sloshing to suppressing runaway cooling of the ICM. 
\end{abstract}


%%%%%%%%%%%%%%%%%%%%%%%%%%%%%%%%%%%%%%%%%%%%%%%%%%%%%%%%%%%%
\section{Introduction}

Galaxy clusters represent the largest known population of gravitationally bound objects formed in the universe. They are still growing in mass today through mergers with small or similar mass objects, as well as through continuous accretion of material from their surrounding large-scale environments. Massive clusters contain a large amount of X-ray emitting hot gas \citep[$\sim 10 - 13$\,\% of total mass; e.g.,][]{Vikhlinin06, Umetsu09, Donahue14} in the intracluster medium (ICM), which is thermalized within the gravitational potential well dominated by dark matter ($\sim 85\%$ in mass). The thermal evolution of the ICM is thus tightly coupled with the evolution of dark matter halos hosting galaxy clusters \citep[e.g.,][]{Lau15, Fujita18}.

Cool cores are often located at the center of galaxy clusters. They are in the form of the dense, relatively cool, metal enriched ICM. The presence of cool cores poses a challenge for our understanding of the thermal evolution of the ICM. In general, X-ray emitting gas is cooling by X-ray emission, and the X-ray luminosity (or its energy loss rate) is proportional to the square of the electron density. Therefore, runaway cooling is expected to take place in the cool cores because the cooling time estimated by the electron density is much shorter than the age of galaxy clusters \citep[for a review, see, e.g.,][]{Peterson06}. Then, cool cores are not expected to be able to survive stably over a cosmological time scale. However, cool cores in galaxy clusters are observed over a wide redshift range beyond $z = 1$ \citep[e.g.,][]{Semler12, Ruppin20}. Moreover, \cite{Tamura01} and \cite{Peterson01} found that the ICM temperature in cool cores remains as hot as several keV, indicating that the ICM in the cool cores is heated. Thus, a thermally stable heating mechanism is required to balance cooling and heating in the cool cores over a cosmological time scale.

Understanding the origin of heating sources and heating mechanisms is one of the challenging and unresolved problems in cluster astrophysics. Feedback from active galactic nuclei (AGNs) in the brightest cluster galaxies (BCGs) is considered to be a plausible mechanism \citep[for reviews, see, e.g.,][]{Fabian12, McNamara12}. The injection of mechanical energy by AGN jets into the ICM (via sound wave heating) is recognized as a heating source. In fact, X-ray cavities are found in a large sample of cool-core clusters and some of them are associated with radio jets \citep[e.g.,][]{McNamara00, McNamara05, Fabian06, Gitti06}. The amount of mechanical power inferred from the size of X-ray cavities appears to be sufficient to keep the balance between cooling and heating \citep[e.g.,][]{Birzan04, Rafferty06, Rafferty08, Hlavacek-Larrondo12, Hlavacek-Larrondo13b, Hlavacek-Larrondo15, Liu19, Prasow-Emond20}.

On the other hand, cluster mergers with a nonzero impact parameter can induce bulk gas motions in cool cores through the transfer of angular momentum \citep[or referred to as gas sloshing;][]{Ascasibar06,Markevitch07}. Thus, gas sloshing has been proposed as an alternative heating source \citep[e.g.,][]{Fujita04b, ZuHone10}. The kinetic energy given by cluster mergers is converted into heat by dissipating turbulence produced by certain mechanisms, such as the Kelvin--Helmholtz instability \citep[e.g.,][]{ZuHone10, Roediger12, ZuHone16, Su17, Ichinohe19, Fujita20}. A characteristic spiral-like feature is often found in the cool cores and is recognized as evidence of gas sloshing \citep[e.g.,][]{Churazov03, Clarke04, Blanton11, Owers11, OSullivan12, Rossetti13, Ghizzardi14, Ichinohe15, Ueda17, Su17, Liu18}, where a merger is expected to take place in the plane of the sky. Moreover, complex X-ray features are identified as evidence of gas sloshing for line-of-sight dark matter mergers \citep[e.g.,][]{Ueda19} or for more complicated merger geometries \citep[e.g.,][]{Ueda20}. 

The main characteristic features of gas sloshing found in galaxy clusters are that (1) the ICM in the cool-core region exhibits a characteristic pair of positive and negative density perturbations in the X-ray brightness distribution, (2) the ICM temperature (metal abundance) in the positive excess regions is lower (higher) than that in the negative excess regions, and (3) the ICM in both excess regions is in pressure equilibrium. However, gas sloshing is not a local mechanism but associated with mergers, so that it is not trivial if gas sloshing can self-regulate cool cores.

Motivated by this, \cite{Ueda20} conducted a systematic study of ICM perturbations in the cool cores for a subsample of 12 cool-core clusters selected from the CLASH \citep{Postman12} sample of 25 massive clusters, for which deep multiwavelength observations are available \citep[see, e.g.,][]{Biviano13, Donahue14, Donahue16, Umetsu14, Umetsu18, Zitrin15, Czakon15}. They developed a detection algorithm for gas density perturbations in cool cores and analyzed X-ray residual image characteristics after subtracting their global profile of the \Chandra\ X-ray brightness distribution, finding that all clusters in their sample have at least a pair of positive and negative excess regions. 

Moreover, \cite{Ueda20} showed that the ICM temperature in the positive excess region of the X-ray residual image is systematically lower than that in the negative excess region and the ICM in both regions is in pressure equilibrium, indicating the ubiquitous presence of gas sloshing features in cool cores. They performed a high-resolution hydrodynamic simulation of a binary cluster merger with 1:3 mass ratio and found that the observed thermodynamic characteristics are consistent with the simulation results, suggesting that the observed gas perturbations can be generated by infalling substructures. However, the analysis of \cite{Ueda20} is limited to a relatively small sample of very high-mass clusters \citep[$M\sim 10^{15}M_{\odot}$;][]{Umetsu16}. Therefore, we need a larger sample of clusters spanning a wider range in mass and redshift to improve the understanding on the origin of heating sources in cluster cores.

In this study, we aim to characterize systematic spatial perturbations in the thermodynamic cool-core properties focusing on the HIghest X-ray FLUx Galaxy Cluster Sample (HIFLUGCS). HIFLUGCS is a sample of 64 galaxy clusters selected from the {\it ROSAT} All-Sky Survey with a flux limit of $2 \times 10^{-11}$\,erg\,s$^{-1}$\,cm$^{-2}$ in the $0.1-2.4$\,keV band and a Galactic latitude limit of $|b| > 20^{\circ}$ covering two-thirds of the sky \citep{Reiprich02}. It includes not only massive galaxy clusters, such as RXC~J1504.1$-$0248, but also nearby galaxy groups and massive elliptical galaxies (central dominant galaxies in groups and clusters), such as MKW3S and NGC~0507. Numerous studies have been made to characterize and understand the statistical properties of the HIFLUGCS sample  \citep[e.g.,][]{Ikebe02, Reiprich06, Chen07, Mittal09, Hudson10, Zhang11, Zhang11b, Lagana11, Mittal11}. \cite{Hudson10} analyzed high-quality X-ray data taken with \Chandra\ and explored cool-core diagnostics in detail. They classified the HIFLUGCS sample into three types using the central cooling time ($t_\mathrm{cool}$), namely, galaxy clusters with  strong cool cores ($t_\mathrm{cool} < 1.0$\,Gyr), weak cool cores ($t_\mathrm{cool} < 7.7$\,Gyr), and non-cool cores. 

For the present study, we have selected from HIFLUGCS a subsample of 28 strong cool-core systems spanning the redshift range $0.0046 \le z \le 0.21530$ and analyzed their archival data taken with \Chandra\ X-ray observations (Table~\ref{tab:list}).  All the strong cool-core systems identified by \citet{Hudson10} have been analyzed in this study. Our sample has a median redshift of $z=0.04195$ and includes nearby galaxy groups and massive galaxy clusters. Hence, this work is complementary to the CLASH analysis presented in \citet{Ueda20} in terms of the mass and redshift range.

The paper is organized as follows. Section~\ref{sec:obs} summarizes the observations and data reduction. Section~\ref{sec:xray} describes our X-ray imaging and spectral analysis of the locally perturbed regions in cluster cores. In Section~\ref{sec:discussion}, we characterize the correlation of thermodynamic properties between the positive and negative excess regions and discuss the results and their implications. Finally, conclusions of this paper are summarized in Section~\ref{sec:conclusions}.

Throughout the paper, we assume a spatially flat $\Lambda$CDM model with a matter density parameter of $\Omega_\mathrm{m}=0.27$ and a Hubble constant of $H_{0} = 70\,$km\,s$^{-1}$\,Mpc$^{-1}$. We use the standard notation $M_{\Delta}$ for the mass enclosed within a sphere of radius $r_{\Delta}$, within which the mean overdensity is $\Delta$ times the critical density of the universe at a particular redshift $z$. Unless otherwise stated, quoted errors are $1\sigma$ uncertainties.

\section{Observations and Data Reduction}
\label{sec:obs}

We analyzed archival X-ray data of each galaxy cluster taken with the Advanced CCD Imaging Spectrometer \citep[ACIS;][]{Garmire03} on board the {\it Chandra X-ray Observatory}. The observation identification (ObsID) numbers of \Chandra\ observations analyzed in this study are summarized in Table~\ref{tab:list}. We used the versions of 4.12 and 4.9.2.1 for \Chandra\ Interactive Analysis of Observations \cite[CIAO;][]{Fruscione06} and the calibration database (CALDB), respectively. We checked the light curve of each dataset using the {\tt lc\_clean} task in CIAO, filtering flare data. The net exposure time of each sample is also shown in Table~\ref{tab:list}. The blanksky data included in the CALDB were adopted as background data. We extracted X-ray spectra from each dataset with {\tt specexctract} in CIAO and combined them after making individual spectrum, response, and ancillary response files for the spectral fitting. We used {\tt XSPEC} version 12.11.0 \citep{Arnaud96} and the atomic database for plasma emission modeling version 3.0.9 in the X-ray spectral analysis, assuming that the ICM is in collisional ionization equilibrium \citep{Smith01}. 

The abundance table of \cite{Lodders09} was used. Here the abundance of a given element is defined as $Z_{i} = (n_{i, {\rm obs}}/n_{\rm H, obs}) / (n_{i, \odot} / n_{\rm H, \odot})$, where $n_{i}$ and $n_{\rm H}$ are the number densities of the $i$th element and hydrogen, respectively. In general, since it is difficult to measure the abundance of individual elements except for iron, we use the iron abundance to represent the ICM metal abundance, such that the abundance of other elements is tied to the iron abundance as $Z_{i} = Z_{\rm Fe}$. The Galactic absorption (i.e., $N_{\rm H}$) for each cluster was estimated using \cite{HI4PI16} and fixed to the estimated value in the X-ray spectral analysis.

%%%%%%%%%%%%%%%%%%%%%%%%%%%%
\begin{table*}[ht]
\begin{center}
\caption{
Summary of our sample objects and \Chandra\ X-ray observations.
}\label{tab:list}
\begin{tabular}{lcccccp{6.5cm}}
\hline\hline	
Cluster			& R.A.\tablenotemark{a}			& Decl.\tablenotemark{a}			& Redshift	& $kT_{\rm vir}$ (keV)\tablenotemark{b}	& $t_\mathrm{exp}$ (ks)\tablenotemark{c}	& ObsID\tablenotemark{d}\\\hline
% & & & & (ks) & \\\hline
A85				& 00:41:50.34	& -09:18:10.91	& 0.05506	& $6.00^{+0.11}_{-0.11}$	& 195.2			& 904, 15173, 15174, 16263, 16264		\\
A133			& 01:02:41.75	& -21:52:55.50	& 0.05660	& $3.96^{+0.08}_{-0.10}$	& 153.3			& 2203, 9897, 13518						\\
NGC~0507			& 01:23:39.90	& 33:15:21.95	& 0.01646	& $1.44^{+0.08}_{-0.10}$	& 58.4			& 317, 2882								\\
A262			& 01:52:46.20	& 36:09:13.08	& 0.01742	& $2.44^{+0.03}_{-0.04}$	& 139.4			& 2215, 7921							\\
A3112			& 03:17:57.67	& -44:14:17.44	& 0.07525	& $4.73^{+0.11}_{-0.12}$	& 133.9			& 2216, 2516, 6972, 7323, 7324, 13135	\\
Fornax cluster	& 03:38:29.04	& -35:27:01.78	& 0.00460	& $1.34^{+0.00}_{-0.00}$	& 366.3			& 239, 319, 2942, 9530, 9798, 9799, 14527, 14529, 16639	\\
2A0335$+$096 		& 03:38:41.06	& 09:58:01.42	& 0.03634	& $3.53^{+0.10}_{-0.13}$	& 103.0			& 919, 7939, 9792						\\
A478			& 04:13:25.15	& 10:27:54.94	& 0.08810	& $7.34^{+0.18}_{-0.19}$	& 50.1			& 1669, 6102							\\
RXJ~0419.6$+$0225	& 04:19:37.94	& 02:24:36.23	& 0.01230	& $1.34^{+0.00}_{-0.00}$	& 108.6			& 3186, 3187, 5800, 5801				\\
EXO~0422$-$086 	& 04:25:51.24	& -08:33:37.32	& 0.03970	& $2.93^{+0.13}_{-0.12}$	& 78.8			& 3970, 4183, 19593, 20862, 20863		\\
A496			& 04:33:37.80	& -13:15:40.05	& 0.03290	& $4.86^{+0.05}_{-0.06}$	& 71.0			& 931, 3361, 4976						\\
Hydra A			& 09:18:05.67	& -12:05:43.95	& 0.05488	& $3.45^{+0.08}_{-0.09}$	& 224.3			& 575, 576, 4969, 4970					\\
MKW4			& 12:04:27.11	& 01:53:45.25	& 0.02000	& $2.01^{+0.04}_{-0.04}$	& 30.0			& 3234									\\
NGC~4636			& 12:42:49.67	& 02:41:12.58	& 0.00313	& $0.90^{+0.02}_{-0.02}$	& 199.2			& 323, 324, 3926, 4415					\\
A3526			& 12:48:48.94	& -41:18:45.34	& 0.01140	& $3.92^{+0.02}_{-0.02}$	& 778.5			& 504, 505, 4190, 4191, 4954, 4955, 5310, 16223, 16224, 16225, 16534, 16607, 16608, 16609, 16610	\\
A1644			& 12:57:11.61	& -17:24:33.09	& 0.04730	& $5.09^{+0.09}_{-0.09}$	& 69.3			& 2206, 7922							\\
NGC~5044			& 13:15:23.99	& -16:23:07.59	& 0.00928	& $1.22^{+0.03}_{-0.04}$	& 419.2			& 798, 9399, 17195, 17196, 17653, 17654, 17666	\\
A1795			& 13:48:52.44	& 26:35:36.00	& 0.06248	& $6.08^{+0.07}_{-0.07}$	& 744.7			& 493, 494, 3666, 5286, 5287, 5288, 5289, 5290, 6159, 6160, 6161, 6162, 6163, 10898, 10899, 10900, 10901, 12026, 12027, 12028, 12029, 13106, 13107, 13108, 13109, 13110, 13111, 13112, 13113, 13412, 13413, 13414, 13415, 13416, 13417, 14268, 14269, 14270, 14271, 14272, 14273, 14274, 14275, 15485, 15486, 15487, 15488, 15489, 17228	\\
A3581			& 14:07:29.79	& -27:01:04.45	& 0.02300	& $1.97^{+0.07}_{-0.07}$	& 91.7			& 1650, 12884							\\
RXC~J1504.1$-$0248	& 15:04:07.48	& -02:48:17.25	& 0.21530	& $9.53^{+1.39}_{-1.16}$	& 161.7			& 4935, 5793, 17197, 17669, 17670		\\
A2029			& 15:10:56.09	& 05:44:41.10	& 0.07728	& $8.26^{+0.09}_{-0.09}$	& 107.6			& 891, 4977, 6101						\\
A2052			& 15:16:43.59	& 07:01:20.25	& 0.03549	& $3.35^{+0.02}_{-0.02}$	& 527.1			& 890, 10477, 10478, 10479, 10480, 10879, 10914, 10915, 10916, 10917	\\
MKW3S			& 15:21:51.86	& 07:42:31.29	& 0.04420	& $3.90^{+0.09}_{-0.09}$	& 57.3			& 900									\\
A2199			& 16:28:38.30	& 39:33:03.84	& 0.03015	& $4.37^{+0.07}_{-0.07}$	& 154.3			& 497, 498, 10748, 10803, 10804, 10805	\\
A2204			& 16:32:46.96	& 05:34:31.95	& 0.15216	& $8.92^{+0.72}_{-0.61}$	& 96.8			& 499, 6104, 7940						\\
AS1101			& 23:13:58.72	& -42:43:31.86	& 0.05800	& $2.57^{+0.12}_{-0.13}$	& 107.7			& 1668, 11758							\\
A2597			& 23:25:19.77	& -12:07:26.04	& 0.08520	& $4.05^{+0.07}_{-0.07}$	& 609.1			& 922, 6934, 7329, 19596, 19597, 19598, 20626, 20627, 20628, 20629, 20805, 20806, 20811, 20817	\\
A4059			& 23:57:00.79	& -34:45:33.44	& 0.04873	& $4.22^{+0.03}_{-0.03}$	& 120.0			& 897, 5785								\\
\hline
\end{tabular}
\end{center}
\tablenotetext{a}{Sky coordinates (J2000.0) of the X-ray brightness peak.}
\tablenotetext{b}{Virial temperature measured by \cite{Hudson10}.}
\tablenotetext{c}{Total next exposure time of \Chandra\ observations.}
\tablenotetext{d}{\Chandra\ observation identification (ObsID) numbers.}
\end{table*}
%%%%%%%%%%%%%%%%%%%%%%%%%%%%

\section{X-ray Imaging and Spectral Analysis of Locally Perturbed Regions}
\label{sec:xray}

\subsection{X-ray imaging analysis}
\label{subsec:ximg}

\begin{figure*}
 \begin{center}
  \includegraphics[width=16cm]{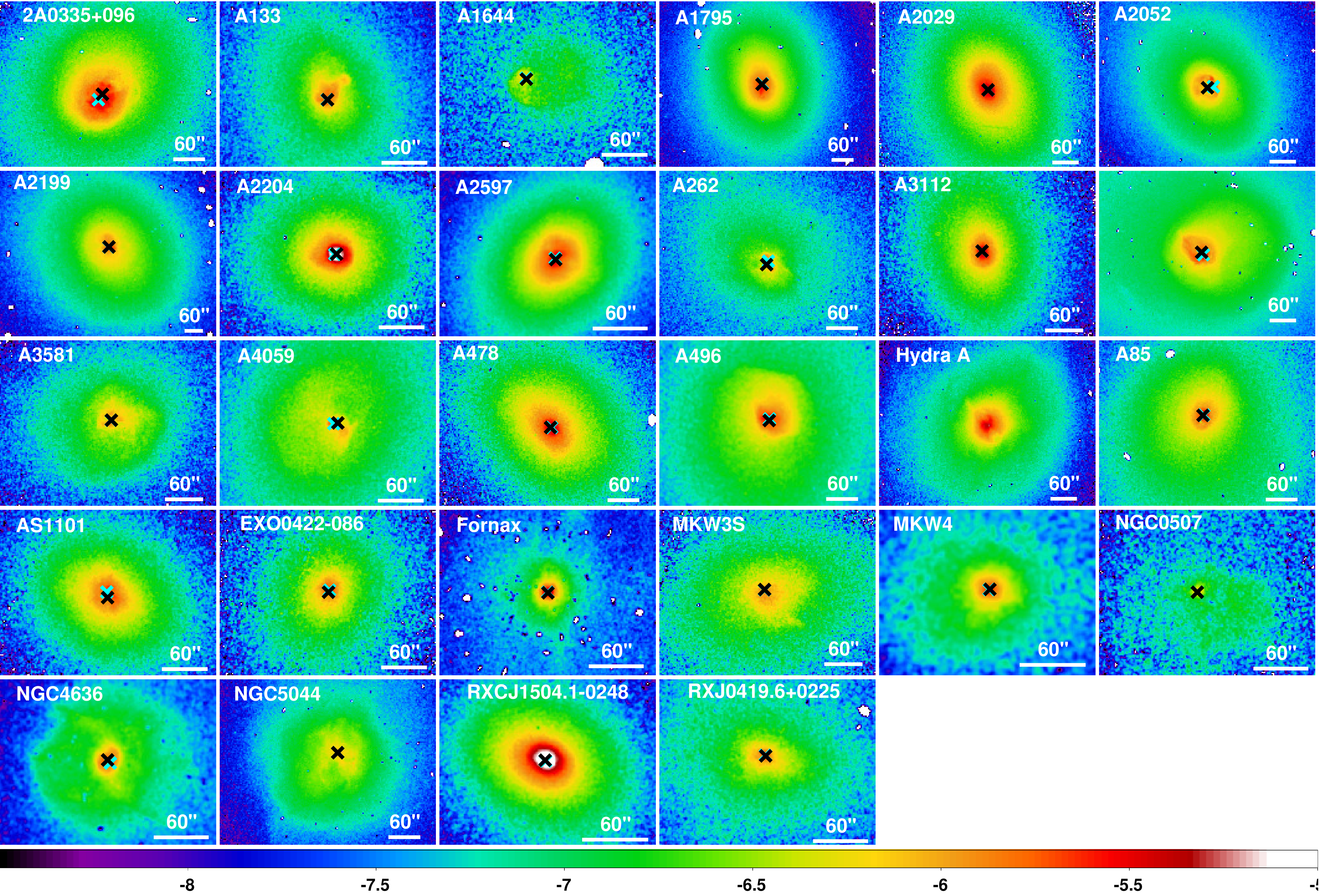}
 \end{center}
\caption{Thumbnails showing the \Chandra\ X-ray brightness distribution in the $0.5$--$7.0$~keV band for 28 systems selected from the HIFLUGS sample.  The color scale is logarithmic and the same in all panels. The color bar indicates the X-ray brightness in units of photon counts\,s$^{-1}$\,arcsec$^{-2}$\,cm$^{-2}$. The images are smoothed with a Gaussian kernel with $2.3\arcsec$ FWHM. The black and cyan crosses in each panel show the positions of the X-ray brightness peak and the BCG, respectively. The white horizontal bar in each panel corresponds to a spatial scale of $60''$. Point sources have been removed.
}
\label{fig:ximg}
\end{figure*}

To detect spatial gas perturbations in the cool core of each system, we analyze their X-ray surface brightness distribution in the $0.5 - 7.0$\,keV band taken with the {\it Chandra X-ray Observatory}. Figure~\ref{fig:ximg} shows the X-ray surface brightness maps of our sample. 

We first determine the mean X-ray brightness distribution using a concentric ellipse fitting algorithm described in \cite{Ueda17} (see also \citealt{Ueda20}).  In our ellipse modeling, the ellipse center is fixed to the peak of the X-ray brightness distribution. This algorithm minimizes the variance of the X-ray surface brightness with respect to the concentric ellipse model. This method has been tested using synthetic X-ray observations of cool cores created from a binary merger simulation \citep{Ueda20}. \cite{Ueda20} found that our detection algorithm can accurately extract the locally perturbed regions in cool cores and can reproduce the temperature difference between the positive and negative excess regions. It was also found that the ellipse model centered on the mass peak and that centered on the X-ray peak perform similarly well for analyses of sloshing cool cores \citep{Ueda20}. We have confirmed this by analyzing independent sets of hydrodynamic simulations of binary cluster mergers \citep{ZuHone10,ZuHone11b,ZuHone16}, as detailed in Section~\ref{sec:comp}, 

We note that since our HIFLUGCS sample includes very nearby objects ($z\ll 0.1$), it is not possible to perform a strictly homogeneous analysis of the whole sample using the same smoothing scale (specified in terms of the  at half maximum (FWHM, hereafter) of the Gaussian kernel) and the same physical image scale. Hence, for clusters at $z < 0.04$ except for A2052, we use different smoothing scales and/or physical image scales for our analysis, as summarized in Table~\ref{tab:pa}. For A2052, its net exposure time exceeds 500\,ks, which is the fourth deepest among our sample, so that the standard parameters are used. 

%%%%%%%%%%%%%%%%%%%%%%%%%%%%
%%%%%%%%%%%%%%%%%%%%%%%%%%%%

%%%%%%%%%%%%%%%%%%%%%%%%%%%%
\begin{figure*}
 \begin{center}
  \includegraphics[width=16cm]{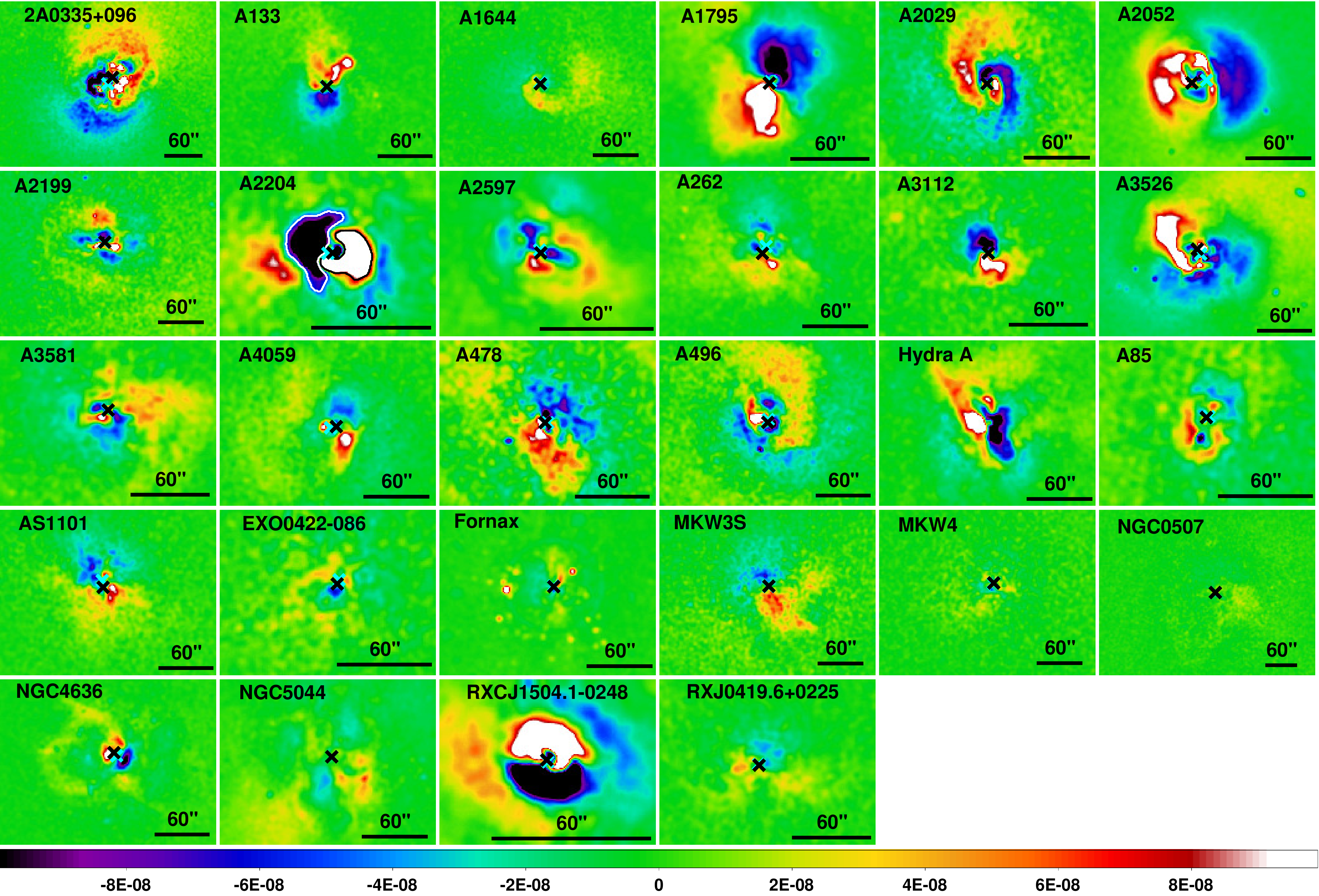}
 \end{center}
\caption{Same as Figure~\ref{fig:ximg} but for the X-ray residual images of our sample. The images are smoothed with a Gaussian kernel with $4.6\arcsec$ FWHM. The color scale is linear and the same in all panels. The black horizontal bar in each panel corresponds to a spatial scale of $60''$. As an example, the identified positive and negative excess regions in A2204 are shown with black and white contours, respectively. 
}
\label{fig:xres}
\end{figure*}
%%%%%%%%%%%%%%%%%%%%%%%%%%%%

The observed position angles and axis ratios of the central X-ray surface brightness for our sample are summarized in Table~\ref{tab:pa}. After ellipse modeling, we subtract off the best-fit model profile from the X-ray brightness distribution. Figure~\ref{fig:xres} shows the X-ray residual images of individual systems in our sample. 

Following \cite{Ueda20}, we define a locally perturbed region as a group of pixels in which the amplitude of a (positive or negative) fluctuation exceeds 20\,\% of the extreme value in the X-ray residual image. If another (unconnected) perturbed region is found, we repeat this procedure using the same threshold value after masking the previous regions. We exclude from our analysis very small (less than 36 pixels) or faint (less than 1000 photon counts) perturbed regions to avoid contamination from point sources. The number of distinct perturbed regions used for our analysis is summarized in Table~\ref{tab:fit}. For each system of our sample, we find at least a pair of positive and negative excess regions in their X-ray residual image. This result indicates that the characteristic dipolar density perturbations in the cool cores exist in not only galaxy clusters but also galaxy groups and massive elliptical galaxies.

%%%%%%%%%%%%%%%%%%%%%%%%%%%%
\begin{table}[ht]
\begin{center}
\caption{
Position angle (PA) and axis ratio (AR) of the central X-ray brightness distribution from ellipse modeling.
}\label{tab:pa}
\begin{tabular}{lcccc}
\hline\hline	
         Cluster &     
         PA\tablenotemark{a} &  
         AR\tablenotemark{b}	& 
         Image size\tablenotemark{c}& 
         FWHM\tablenotemark{d}	\\ 
         		&				(deg)&						& (kpc)			& ($\arcsec$)				\\	\hline
             A85 &  66 & 0.84	& 200 & 2.3	\\
            A133 &  88 & 0.80 	& 200 & 2.3		\\
        NGC~0507 &  90 & 0.87	& 100 & 2.3 	\\
            A262 & 114 & 0.70	& 100 & 2.3 	\\
           A3112 & 100 & 0.75	& 200 & 2.3 	\\
  Fornax cluster &   0 & 0.94	&  50 & 2.3 	\\
    2A0335$+$096 &  49 & 0.88	& 200 & 3.5  	\\
            A478 & 132 & 0.73	& 200 & 2.3 	\\
RXJ~0419.6$+$0225 & 165 & 0.81	& 100 & 2.3  	\\
  EXO~0422$-$086 &  71 & 0.85	& 200 & 2.3  	\\
            A496 & 113 & 0.96	& 200 & 4.6 	\\
         Hydra A &  45 & 0.76	& 200 & 2.3 	\\
            MKW4 &   5 & 0.92	& 100 & 4.6 	\\
        NGC~4636 &  75 & 0.89	& 200 & 3.5 	\\
           A3526 &  18 & 0.92 	& 100 & 2.3	\\
           A1644 & 178 & 0.81	& 200 & 2.3   \\
        NGC~5044 &  88 & 0.95	&  50 & 2.3 	\\
           A1795 &  96 & 0.74 	& 200 & 2.3 \\
           A3581 & 160 & 1.00	& 100 & 2.3  \\
RXC~J1504.1$-$0248 & 156 & 0.78	& 200 & 2.3  \\
           A2029 & 113 & 0.79 	& 200 & 2.3 \\
           A2052 & 144 & 0.86 	& 200 & 2.3 \\
           MKW3S &   5 & 0.79	& 200 & 2.3  \\
           A2199 & 123 & 0.83	& 200 & 4.6  	\\
           A2204 & 150 & 0.92	& 200 & 2.3  \\
          AS1101 & 142 & 0.80	& 200 & 2.3   \\
           A2597 &  52 & 0.77	& 200 & 2.3   \\
           A4059 &  70 & 0.88  	& 200 & 2.3 \\
\hline
\end{tabular}
\end{center}
\tablenotetext{a}{Position angle measured south of east. The typical uncertainty is $\pm 1^{\circ}$.}
\tablenotetext{b}{Axis ratio ($\le 1$). The typical uncertainty  is $\pm 0.01$.}
\tablenotetext{c}{Physical size of the X-ray image used for the analysis.}
\tablenotetext{d}{FWHM of the Gaussian smoothing kernel.}
\end{table}
%%%%%%%%%%%%%%%%%%%%%%%%%%%%

\begin{figure}
 \begin{center}
  \includegraphics[scale=0.28,clip]{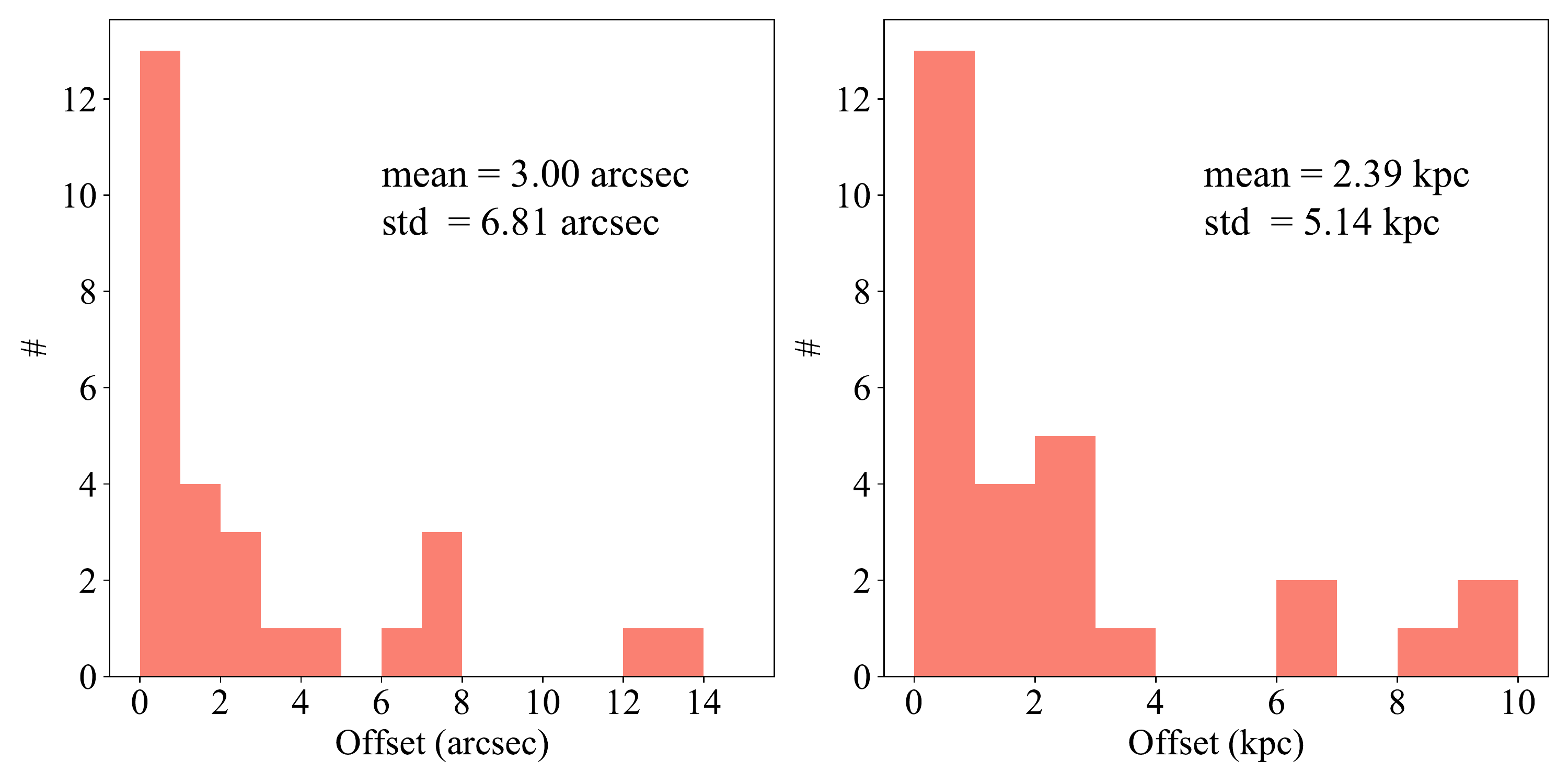}
 \end{center}
\caption{Distribution of positional offsets between the X-ray brighness peak and the BCG position for each system, shown in units of arcseconds (left) or kpc (right). The values of the absolute mean offset $\langle|\mbox{\boldmath{$d$}}|\rangle$ and the standard deviation $\sigma(\mbox{\boldmath{$d$}})$ are listed in each panel. 
}
\label{fig:offset}
\end{figure}

Figure~\ref{fig:offset} shows the distribution of positional offsets between the X-ray brightness peak and the BCG position. We find that the typical positional offset is less than 3\,kpc, so that the BCG and the X-ray peak are aligned well with each other. We note that the sloshing features in NGC~5044 were observed and reported by \cite{Gastaldello13}.

%%%%%%%%%%%%%%%%%%%%%%%%%%%%
%%%%%%%%%%%%%%%%%%%%%%%%%%%%

\subsection{X-ray spectral analysis}
\label{sec:spec}

We extract an X-ray spectrum of each perturbed region identified in Section~\ref{subsec:ximg}. The X-ray spectra of the ICM in the $0.4$--$7.0$~keV band are analyzed using the model of {\tt phabs * apec} in {\tt XSPEC}. The cluster redshift is fixed to the value listed in Table~\ref{tab:list}. The best-fit parameters of the ICM temperature and abundance in the perturbed regions are listed in Table~\ref{tab:fit}. Thanks to the high-quality data taken with \Chandra, both parameters are well determined for all individual objects in our sample. We also measure the ICM electron number density ($n_{\rm e}$), pressure ($p_{\rm} = kT \times n_{\rm e}$), and entropy ($K_{\rm e} = kT \times n_{\rm e}^{-2/3}$), assuming a line-of-sight length of $L=1$~Mpc. Since the observed X-ray surface brightness scales as $S_\mathrm{X}\propto n_\mathrm{e}^2L$, $n_\mathrm{e}$ scales as  ($L$/1\,{\rm Mpc})$^{-1/2}$. Table~\ref{tab:npe} summarizes these thermodynamic parameters measured in the positive and negative excess regions. It should be noted that all observed quantities are projected ones (i.e., not deprojected).

\begin{figure*}
 \begin{center}
  \includegraphics[scale=0.32,clip]{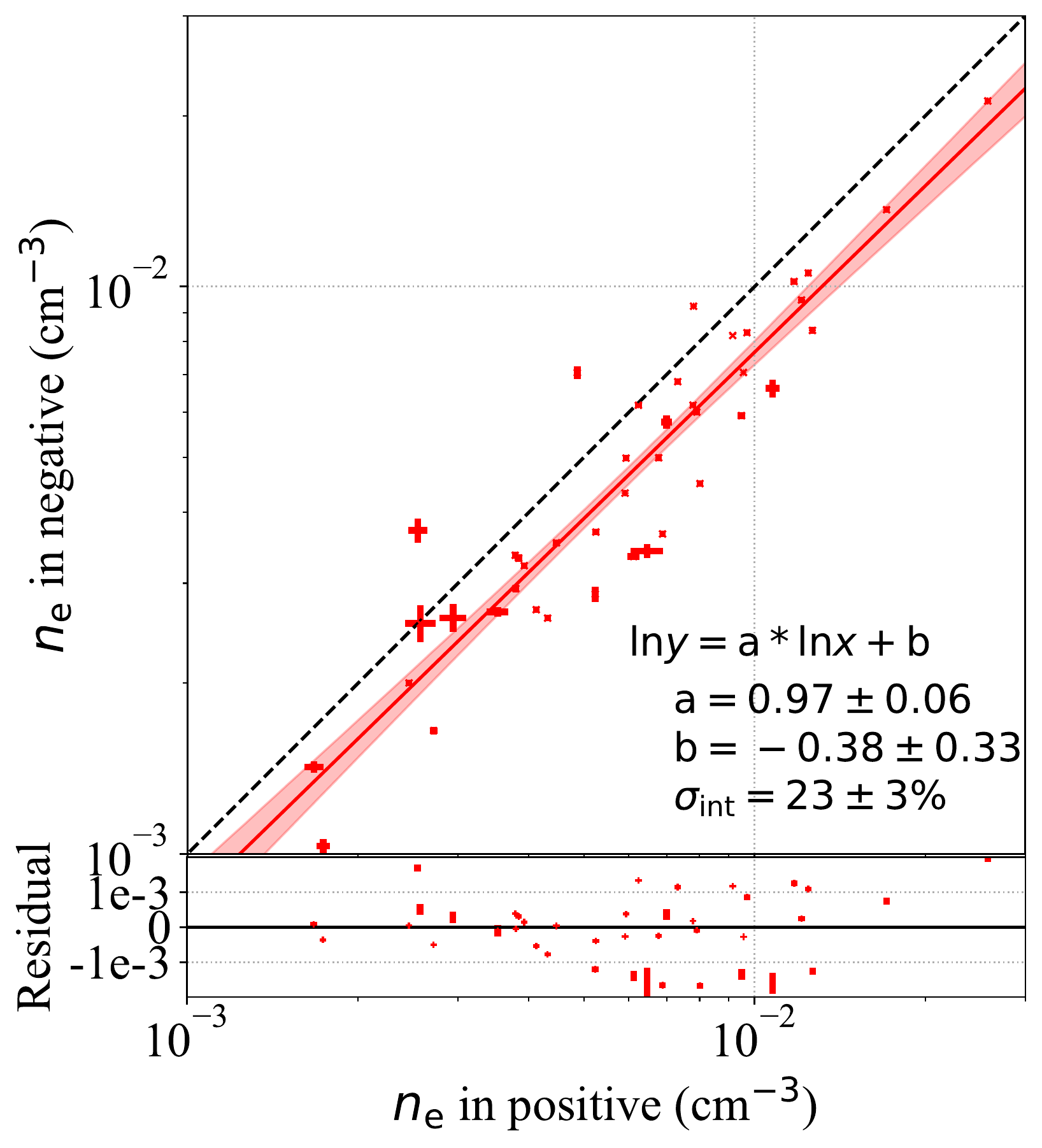}
  \includegraphics[scale=0.32,clip]{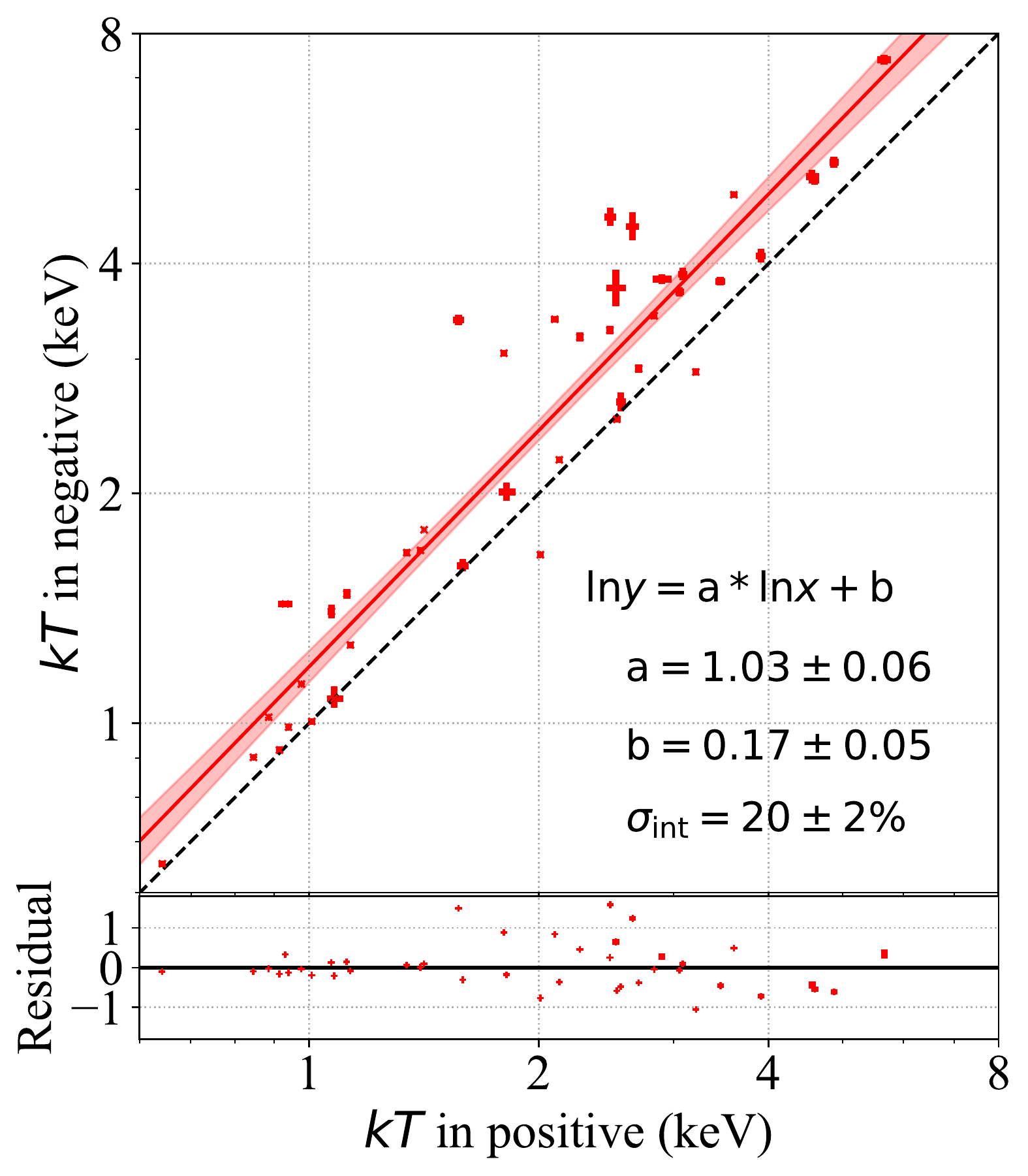}
  \includegraphics[scale=0.32,clip]{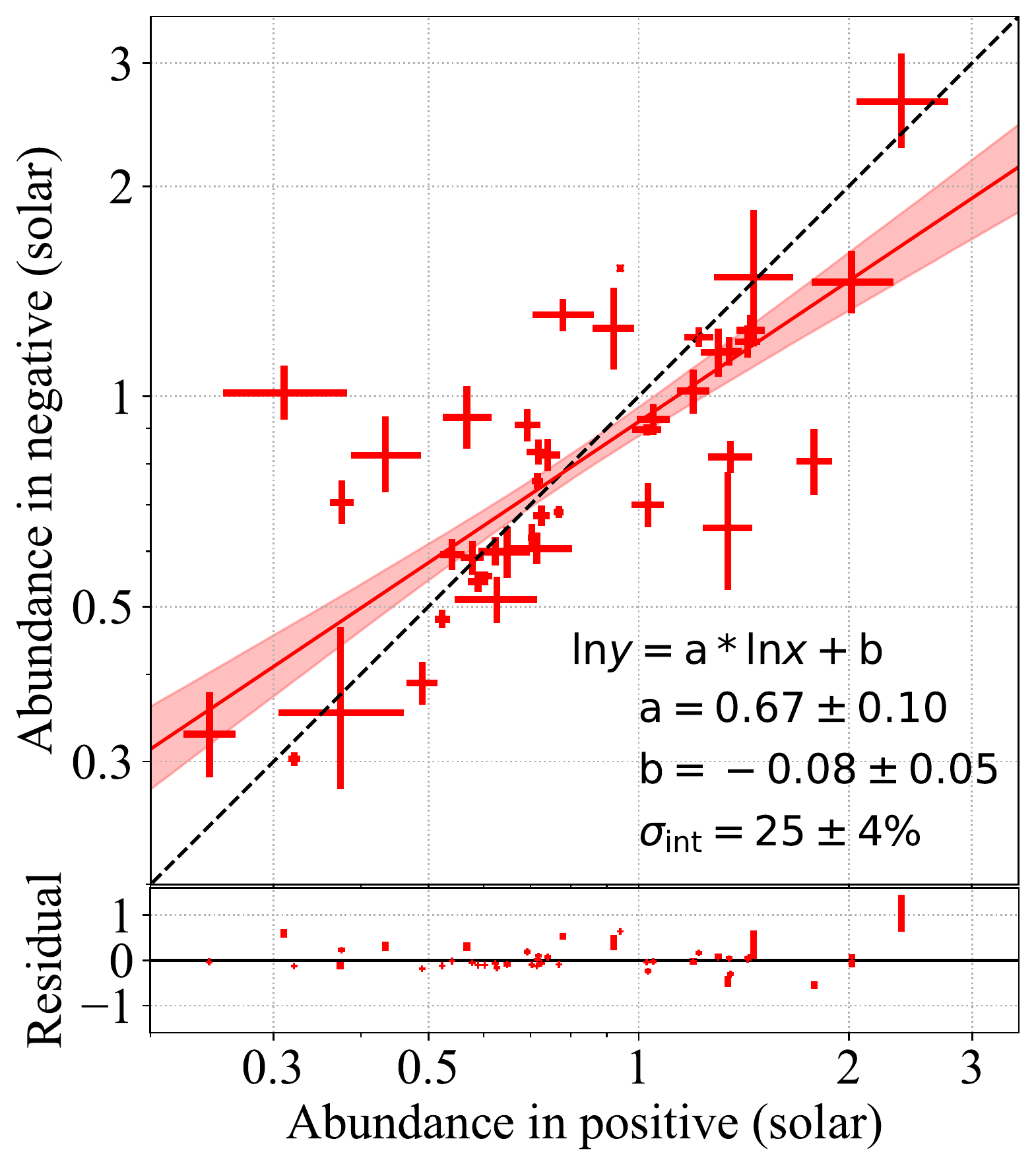}
  \includegraphics[scale=0.32,clip]{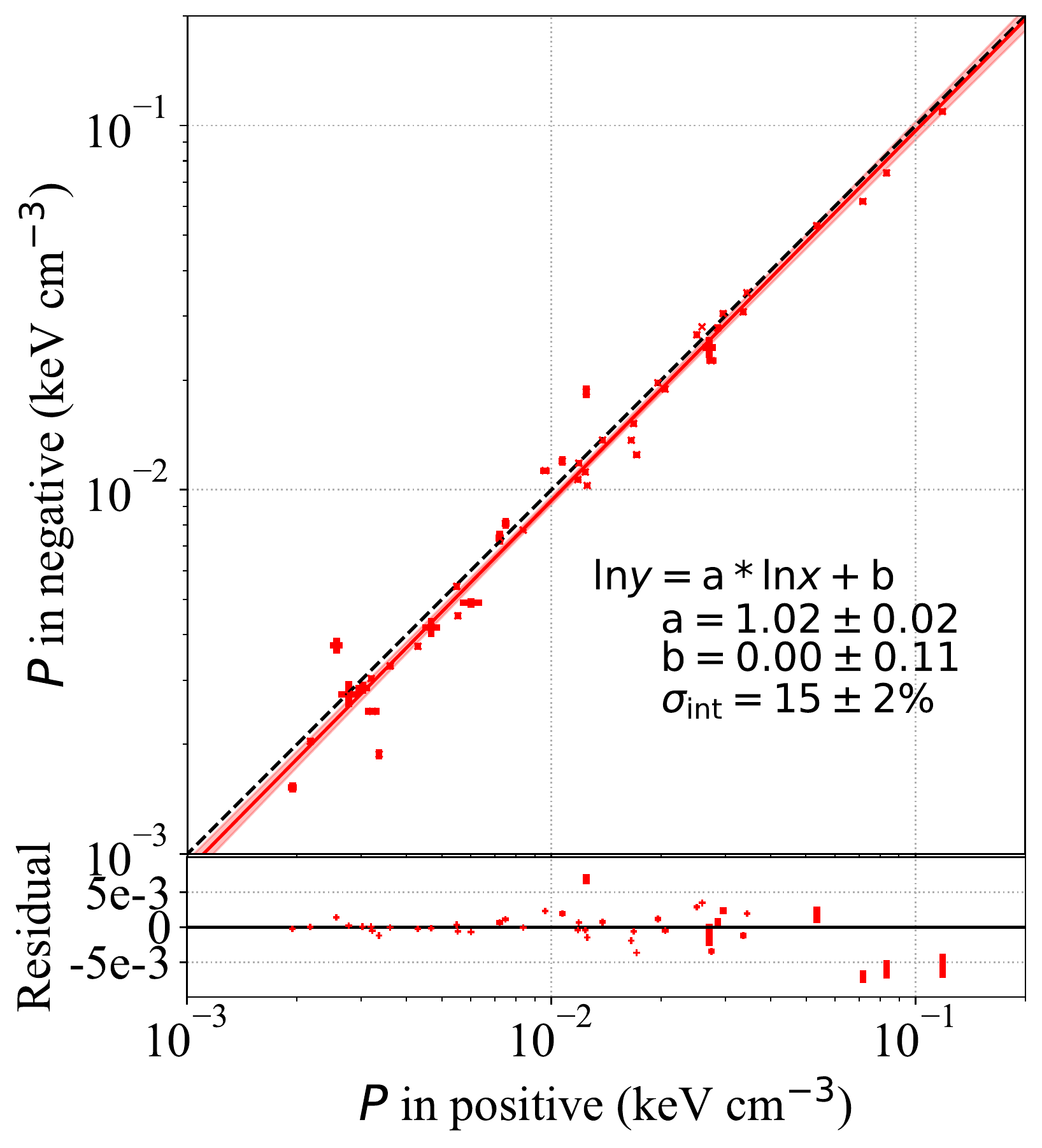}
  \includegraphics[scale=0.32,clip]{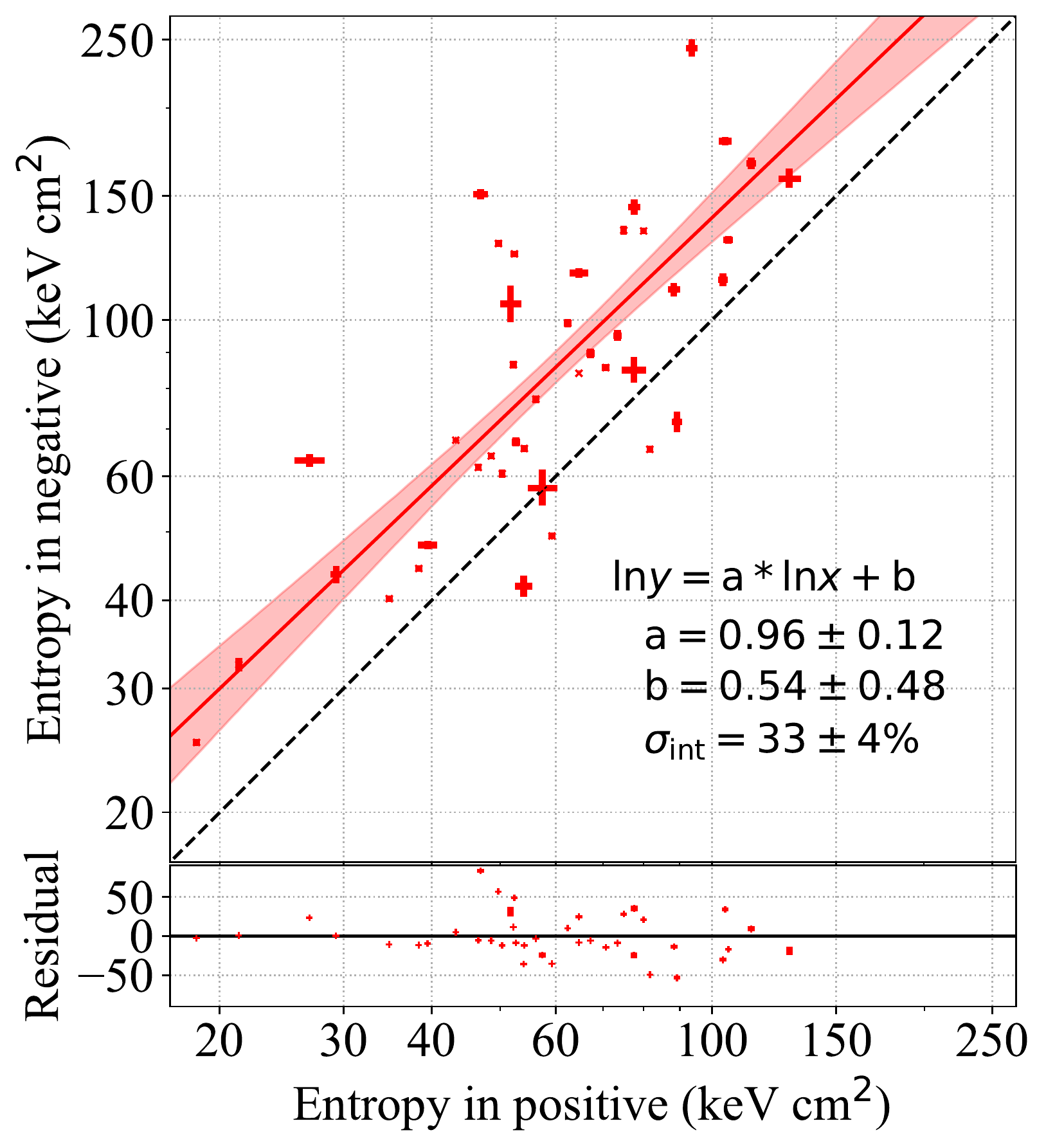}
 \end{center}
\caption{Comparison of the ICM properties measured in the positive and negative excess regions for our sample: electron number density (top left), temperature (top middle), metal abundance (top right), pressure (bottom left), and entropy (bottom right). Our measurements are shown with red error bars. In each panel, the black dashed line shows the one-to-one relation and the red line shows the best-fit regression model (see Table~\ref{tab:MCMC}). The red shaded region indicates the 68\% confidence interval around the best-fit relation. Data with no counterpart region are excluded. The best-fit residuals are shown in the lower panel of each plot.
}
\label{fig:para}
\end{figure*}

We find that the ICM temperature (electron number density) in the positive excess regions is systematically lower (higher) than that in the negative excess regions. Hence, the difference in entropy between the two regions is further enhanced, whereas the electron pressure in the two regions is less perturbed. Figure~\ref{fig:para} compares the ICM properties between the positive and negative excess regions for our sample.  In Section~\ref{sec:thermo}, we will characterize and discuss in detail the correlation for each thermodynamic quantity.

%%%%%%%%%%%%%%%%%%%%%%%%%%%%
\input{table/table_kTa.tex}
\input{table/table_kTb.tex}
%%%%%%%%%%%%%%%%%%%%%%%%%%%%

%%%%%%%%%%%%%%%%%%%%%%%%%%%%
\input{table/table_nPKa.tex}
\input{table/table_nPKb.tex}
%%%%%%%%%%%%%%%%%%%%%%%%%%%%

\subsection{Measurements of the X-ray brightness contrast}
\label{sec:co}

For each system in our sample, we measure the X-ray brightness contrast of fluctuations in the positive and negative excess regions defined as $|\Delta I_\mathrm{X}| / \langle I_\mathrm{X} \rangle$. Here $|\Delta I_\mathrm{X}|$ is the difference between the mean values of the X-ray surface brightness in the positive and negative excess regions \citep{Ueda18,Ueda20}, and $\langle I_\mathrm{X} \rangle$ represents the mean X-ray surface brightness averaged within the whole perturbed region of each system. The measurements of the X-ray brightness contrast are summarized in Table~\ref{tab:dIx}. 

%%%%%%%%%%%%%%%%%%%%%%%%%%%%
\begin{figure}
 \begin{center}
  \includegraphics[scale=0.33, clip]{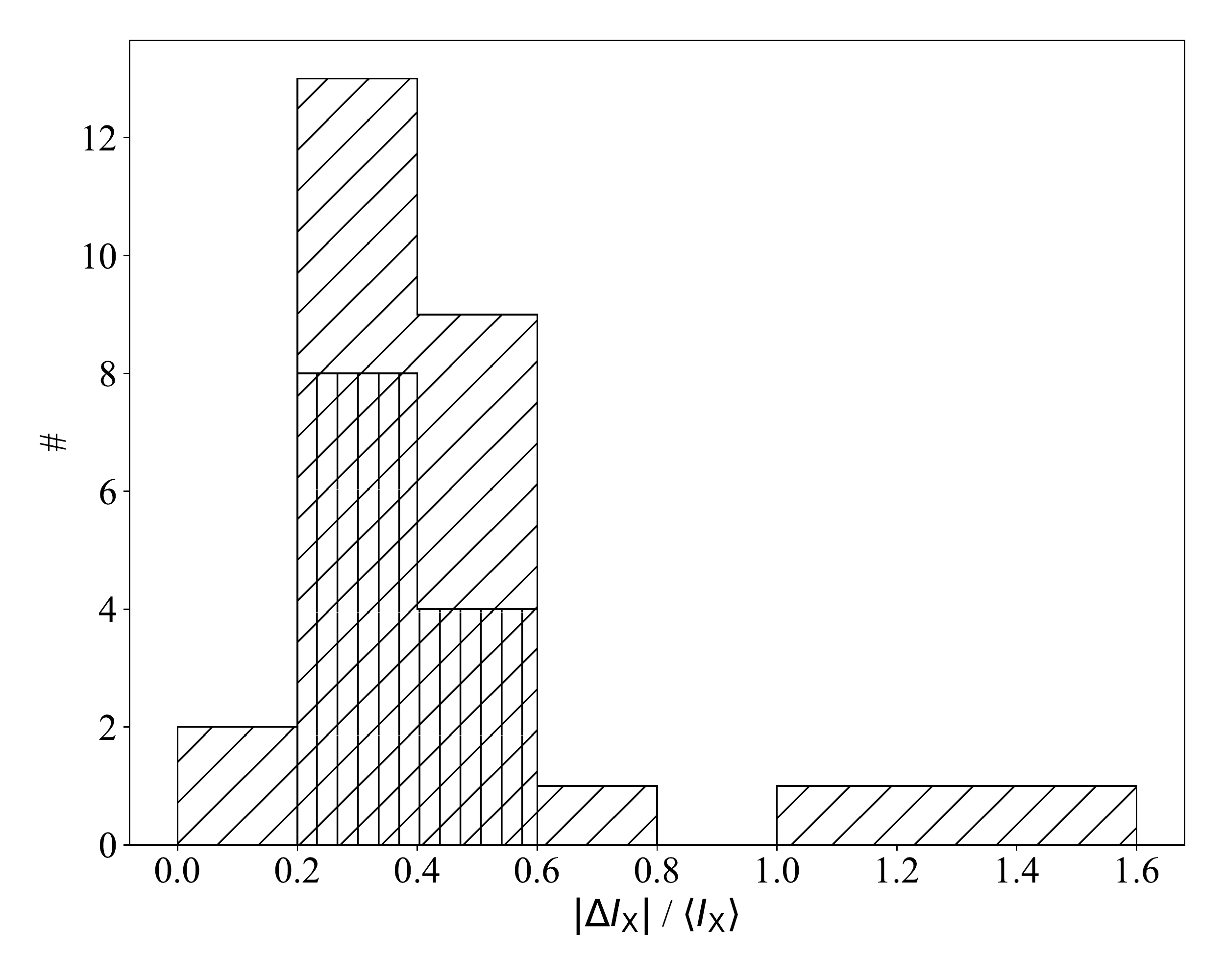}
 \end{center}
\caption{Histogram distribution of the X-ray brightness contrast $|\Delta I_\mathrm{X}| / \langle I_\mathrm{X} \rangle$ in perturbed cool cores. The slash-hatched and vertical-hatched histograms correspond to the HIFLUGCS and CLASH cool-core samples, respectively.
}
\label{fig:histo}
\end{figure}
%%%%%%%%%%%%%%%%%%%%%%%%%%%%

Figure~\ref{fig:histo} shows the histogram distribution of the brightness contrast $|\Delta I_\mathrm{X}|/\langle I_\mathrm{X}\rangle$ for our sample along with the results for the CLASH cool-core sample \citep{Ueda20}. Except for NGC~0507, MKW4, and A1644, the observed values of $|\Delta I_\mathrm{X}|/\langle I_\mathrm{X}\rangle$ are smaller than unity, and the distributions for the HIFLUGCS and CLASH samples are fairly similar.

We note that two of the outliers having $|\Delta I_\mathrm{X}|/\langle I_\mathrm{X}\rangle > 1$ are less massive, galaxy--group scale objects. NGC~0507 ($z=0.01646$) is the brightest member galaxy of a nearby galaxy group. MKW4 is a nearby galaxy group at $z=0.02000$ and its total hydrostatic mass was estimated as $M_{200}=(9.7 \pm 1.5)\times 10^{13}M_{\odot}$ using X-ray observations \citep[][]{Sarkar21}. On the other hand, A1644 ($z=0.04730$) is a nearby cluster containing multiple substructures. \citet{Monteiro-Oliveira20} obtained a weak-lensing mass estimate of $M_{200}=1.90^{+0.89}_{-1.28}\times 10^{14}M_\odot$ for the main cluster component. The cluster is known for a remarkable spiral-like X-ray morphology in the central core region, which is likely due to the result of a collision with a northern substructure \citep{Monteiro-Oliveira20}. The large-scale spiral feature in the X-ray emission is directly visible without subtracting the global brightness profile \citep[][see also Figure~\ref{fig:ximg}]{Johnson10, Lagana10}.

%%%%%%%%%%%%%%%%%%%%%%%%%%%%
\begin{table}[ht]
\begin{center}
\caption{
X-ray brightness contrast in the perturbed regions of each system.
}\label{tab:dIx}
\begin{tabular}{lc}
\hline\hline	
Sample				& $|\Delta I_{\rm X}| / \langle I_{\rm X} \rangle$	\\ \hline
       	    A85  & $0.265 \pm 0.004$ \\
            A133 & $0.458 \pm 0.003$ \\
        NGC~0507 & $1.509 \pm 0.030$ \\
            A262 & $0.442 \pm 0.003$ \\
           A3112 & $0.244 \pm 0.005$ \\
  Fornax cluster & $0.226 \pm 0.005$ \\
    2A0335$+$096 & $0.361 \pm 0.002$ \\
            A478 & $0.325 \pm 0.003$ \\
RXJ~0419.6$+$0225 & $0.485 \pm 0.006$ \\
EXO~0422$-$086 & $0.578 \pm 0.014$ \\
            A496 & $0.404 \pm 0.002$ \\
         Hydra A & $0.228 \pm 0.003$ \\
            MKW4 & $1.170 \pm 0.040$ \\
        NGC~4636 & $0.406 \pm 0.005$ \\
           A3526 & $0.342 \pm 0.002$ \\
           A1644 & $1.308 \pm 0.016$ \\
        NGC~5044 & $0.388 \pm 0.001$ \\	   
           A1795 & $0.238 \pm 0.002$ \\
           A3581 & $0.525 \pm 0.004$ \\
RXC~J1504.1$-$0248 & $0.178 \pm 0.005$ \\
           A2029 & $0.232 \pm 0.002$ \\
           A2052 & $0.436 \pm 0.002$ \\
           MKW3S & $0.684 \pm 0.005$ \\	   
           A2199 & $0.289 \pm 0.003$ \\
           A2204 & $0.243 \pm 0.005$ \\
          AS1101 & $0.388 \pm 0.002$ \\
	  	   A2597 & $0.168 \pm 0.003$ \\
           A4059 & $0.479 \pm 0.003$ \\
\hline
		Mean & 0.464 \\
		Standard deviation	& 0.327 \\
\hline
\end{tabular}
\end{center}
\end{table}
%%%%%%%%%%%%%%%%%%%%%%%%%%%%

%%%%%%%%%%%%%%%%%%%%%%%%%%%%
\begin{table}[ht]
\begin{center}
\caption{
Best-fit regression parameters of the scaling relation 
$\ln{Y} = a\ln{X} + b\pm \sigma_\mathrm{int}$ 
derived for each thermodynamic quantity between the positive ($X$) and negative ($Y$) excess regions in cool cores. In the scaling relations, number densities, temperatures, metal abundances, pressures, and entropies are expressed in units of cm$^{-3}$, keV, $Z_\odot$, keV\,cm$^{-3}$, and keV\,cm$^{2}$, respectively.
}\label{tab:MCMC}
\begin{tabular}{p{2.5cm}ccc}
\hline\hline	
Variable	& Slope ($a$)		& Intercept ($b$)	& $\sigma_{\mathrm{int}}$	\\ \hline
Number density  & $0.97 \pm 0.06$   & $-0.38 \pm 0.33$  & $23 \pm 3$\%  \\
Temperature		& $1.03 \pm 0.06$	& $0.17 \pm 0.05$	& $20 \pm 2$\%								\\
Abundance		& $0.67 \pm 0.10$	& $-0.08 \pm 0.05$	& $25 \pm 4$\%								\\
Pressure		& $1.02 \pm 0.02$	& $0.00 \pm 0.11$	& $15 \pm 2$\%								\\
Entropy			& $0.96 \pm 0.12$	& $0.54 \pm 0.48$	& $33 \pm 4$\%								\\ \hline
Pressure for HIFLUGCS + CLASH\tablenotemark{a}	& $1.01 \pm 0.02$	& $-0.02 \pm 0.08$	& $14 \pm 2$\%	\\
\hline
\end{tabular}
\end{center}
\tablenotetext{a}{\citet{Ueda20}
}
\end{table}
%%%%%%%%%%%%%%%%%%%%%%%%%%%%

%%%%%%%%%%%%%%%%%%%%%%%%%%%%
%%%%%%%%%%%%%%%%%%%%%%%%%%%%

%%%%%%%%%%%%%%%%%%%%%%%%%%%%
\begin{figure}
 \begin{center}
  \includegraphics[scale=0.45,clip]{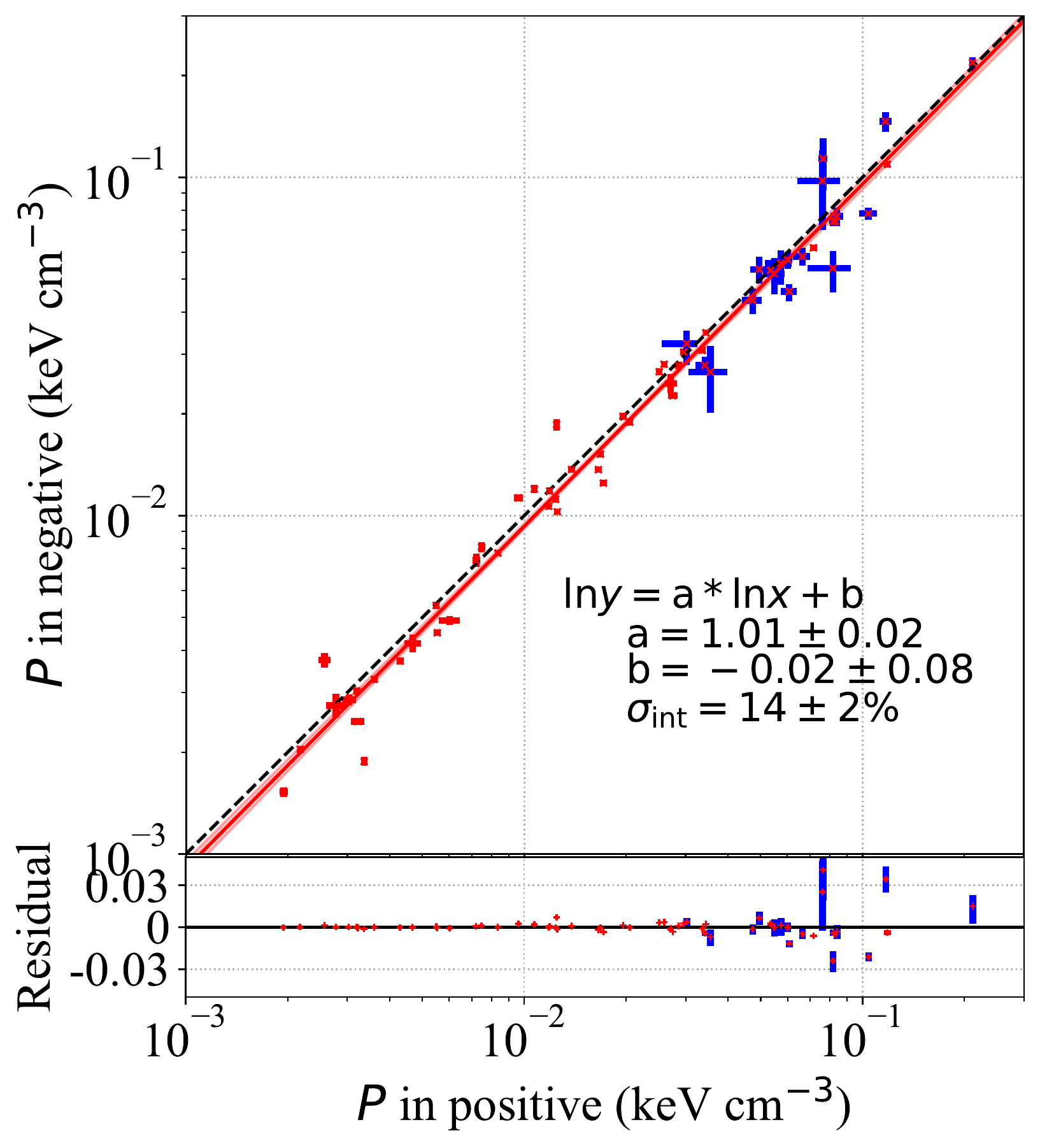}
 \end{center}
\caption{Comparison of the pressures measured in the positive and negative excess regions of cool cores. The measurements for our sample are shown with red error bars (same as those shown in the bottom left panel of Figure~\ref{fig:para}), along with the measurements for the CLASH cool-core sample \citep[blue error bars;][]{Ueda20}. The red solid line shows the the best-fit relation for the composite (HIFLUGCS + CLASH) sample.  The red shaded region indicates the 68\% confidence interval around the best-fit relation. The best-fit residuals are shown in the lower panel.
}
\label{fig:p_wC}
\end{figure}
%%%%%%%%%%%%%%%%%%%%%%%%%%%%

\section{Results and Discussion}
\label{sec:discussion}

\subsection{Thermodynamic properties of the ICM in the perturbed regions}
\label{sec:thermo}

We measured the thermodynamic properties of the perturbed ICM regions in cool cores of our sample selected from HIFLUGCS. Here we compare the ICM properties between the positive and negative excess regions to explore possible origins of the systematic gas perturbations in the cool cores and to understand their characteristics in a statistical manner. Figure~\ref{fig:para} compares the ICM properties (temperature, metal abundance, pressure, and entropy) between the positive and negative excess regions. Since some of the detected perturbed regions have no counterpart detected (e.g., the region \#3 of A262 and the region \#2 of EXO~0422$-$086), we excluded these regions from the comparison. 

To study and characterize the correlation of the ICM properties between the positive and negative excess regions, we perform a linear regression analysis using affine-invariant Markov Chain Monte Carlo sampling \citep{Goodman10} implemented by the  {\tt emcee} python package \citep{Foreman-Mackey13}. Here we regard the ICM properties in the positive excess regions as an independent variable ($x$) for our regression analysis. This is because the ICM in the positive excess regions is expected to be originated from the ICM of cool cores, while the ICM in the negative excess regions from the ambient hot gas of the ICM. In this context, the ICM properties in the positive excess regions serve as a more fundamental quantity for a statistical comparison of cool-core properties in both regions. Our regression model is $y = ax + b$ with $x=\ln{X}$ and $y=\ln{Y}$ (i.e., $Y=e^b X^a$), where $X$ and $Y$ denote the values of each physical quantity in the positive and negative excess regions, respectively (see Tables~\ref{tab:fit} and \ref{tab:npe}).

The log-likelihood function is then written as \citep[see, e.g.,][]{Tian20}
\begin{equation}
-2 \ln \mathcal{L} = \sum_{i} \ln (2 \pi \sigma_{i}^{2}) + \sum_{i} \frac{[y_{i} - (ax_{i} + b)]^{2}}{\sigma_{i}^{2}},
\end{equation}
where $i$ runs over all objects in our sample and $\sigma_{i}$ includes the observational uncertainties in $x$ and $y$, $\sigma_{x_i}$ and $\sigma_{y_i}$,
and lognormal intrinsic scatter, $\sigma_{\rm int}$,
\begin{equation}
\sigma_{i}^{2} = \sigma_{y_{i}}^2 + a^2 \sigma_{x_{i}}^2 + \sigma_{\rm int}^2.
\end{equation}
The $\sigma_\mathrm{int}$ parameter is responsible for the intrinsic scatter around the mean relation due to unaccounted errors and/or astrophysics.
We use uninformative uniform priors on $a$ and $b$ of $a \in [-5, 5]$ and $b \in [-5, 5]$. For the intrinsic scatter, we assume a prior that is uniform in $\ln \sigma_{\rm int}$ in the range $\ln \sigma_{\rm int} \in [-5, 5]$. 
We sample the posterior probability distributions of the regression parameters ($a, b, \sigma_{\rm int}$) over the full parameter space allowed by the priors. Here the errors quoted denote the 16th and 84th percentiles of the marginalized distributions.

For each ICM quantity, the best-fit relation with the marginalized uncertainty is shown in Figure~\ref{fig:para}. The best-fit regression parameters are summarized in Table~\ref{tab:MCMC}. The best-fit relations for number density, temperature, and entropy have slopes consistent with unity while exhibiting a clear offset from the one-to-one relation (see Figure~\ref{fig:para}). For electron number density, temperature, and entropy, we repeated our regression analysis by fixing the respective slopes to unity ($a=1$), finding ensemble averaged ratios of
$n_{\mathrm{e,neg}} / n_{\mathrm{e,pos}} = 0.77^{+0.03}_{-0.02}$,
$T_\mathrm{neg}/T_\mathrm{pos} = 1.20^{+0.04}_{-0.03}$, and 
$K_\mathrm{neg}/K_\mathrm{pos} = 1.43\pm 0.07$.
We find that the best-fit relation for pressure is consistent with the one-to-one relation ($P_\mathrm{neg}=P_\mathrm{pos}$), indicating that the perturbed regions are in pressure equilibrium. On the other hand, the best-fit slope for the ICM metal abundance ($a=0.67\pm 0.10$) is significantly smaller than unity.

A systematic difference in temperature between the positive and negative excess regions is a characteristic feature of gas sloshing. The observed clear offset in temperature is consistent with the picture of gas sloshing. \cite{Ueda20} characterized the temperature difference in the perturbed cool-core regions for the CLASH sample, finding a temperature ratio of $T_\mathrm{neg}/T_\mathrm{pos} = 1.18 \pm 0.05$\footnote{\cite{Ueda20} assumed a linear relation without intercept, i.e., $Y \propto X$, so that their regression model is written as $y = x + b$ with $y=\ln{Y}$ and $x=\ln{X}$. \cite{Ueda20} did not account for the intrinsic scatter around the mean relation.}, which is also consistent with the result of their cluster merger simulation. The observed offset in entropy is also consistent with that for the CLASH cool-core sample, $K_\mathrm{neg}/K_\mathrm{pos} = 1.38 \pm 0.09$ \citep{Ueda20}. Our best-fit relations for both temperature and entropy agree well with the CLASH results. This agreement suggests that the characteristic features in gas sloshing do not depend strongly on the redshift and mass of clusters.

On the other hand, the best-fit slope for the metal abundance relation is smaller than unity, so that the abundance ratio $Z_\mathrm{neg}/Z_\mathrm{pos}$ decreases with increasing $Z_\mathrm{pos}$. In general, the hot gas in the positive excess regions is expected to have a higher ICM abundance because it originates from the ICM in cool cores stripped by gas sloshing \citep[e.g.,][]{Sanders09, Blanton11, Paterno-Mahler13, Ghizzardi14}. However, such a trend in metal abundance is not found in some clusters \citep{Clarke04, Sanders14, Ueda17}. \cite{Ueda20} also found no significant difference in the ICM metal abundance between the positive and negative excess regions (i.e., $Z_\mathrm{neg}\sim Z_\mathrm{pos}$). We will discuss this in more detail in Section~\ref{sec:abund}.

The best-fit pressure relation indicates that the perturbed regions are in pressure equilibrium, which is consistent with the scenario where subsonic gas sloshing generates gas density perturbations in cool cores. To improve the statistics, we combine our sample with the CLASH cool-core sample analyzed by \citet{Ueda20}. We note that the same sample definition and detection algorithm are used in this work and \citet{Ueda20}. Since the CLASH sample is composed of hot ($T_\mathrm{X}>5$~keV), high-mass clusters, the inclusion of the CLASH sample allows us to probe the higher-pressure regime. The composite sample spans a mass range of $M_{500}\in [10^{13}$, $10^{15}] M_\odot$ \citep[][]{Umetsu16,Schellenberger17}.

In Figure~\ref{fig:p_wC}, we compare the ICM pressures measured in the positive and negative excess regions. The best-fit $P_\mathrm{neg}$--$P_\mathrm{pos}$ relation is also shown in the figure. We find that the best-fit relation is remarkably consistent with the one-to-one relation with an intrinsic scatter of $14\pm 2\%$ (Table~\ref{tab:MCMC}). This tight equilibrium relation supports that the origin of these isobaric perturbations can be gas sloshing. This is also in line with the \textit{Hitomi} X-ray measurements of cluster gas motions in the central core of the Perseus cluster \citep{Hitomi16, Hitomi18d}. The observed velocity dispersion including both turbulent and small-scale bulk motions is significantly smaller than the adiabatic sound speed of the ICM.

Since the cool cores in our sample are dominated by isobaric perturbations, the amplitude of entropy fluctuations can be used to constrain the one-dimensional Mach number of turbulence. Following \cite{Gaspari14} and \cite{Hofmann16}, if the ensemble averaged ratio of entropy $K_\mathrm{neg} / K_\mathrm{pos} = 1.43 \pm 0.07$ can be used to represent the amplitude of entropy perturbations, we find a one-dimensional turbulent Mach number of $0.18 \pm 0.03$, corresponding to the three-dimensional Mach number of $0.31 \pm 0.05$. This inferred Mach number is consistent with the {\it Hitomi} observations \citep{Hitomi18d}.

Although the best-fit pressure relation indicates that isobaric perturbations dominate in the cool cores, there is a modest level of intrinsic scatter ($15\pm 2$~\%). Some additional physical processes such as AGN feedback may contribute to the scatter. For example, {\it Chandra} X-ray observations of A2052 exhibit complex substructures in its cool core \citep[e.g.,][]{Blanton03, Blanton11}. For this cluster, in fact, the measured temperatures of the perturbed region \#\,1 deviate significantly from the best-fit temperature relation, suggesting that such additional processes are likely at play in this region. However, we emphasize that the inferred intrinsic scatter in the pressure relation is significantly smaller than that of both temperature and number density (see Table~\ref{tab:MCMC}), indicating that, on average, isobaric perturbations dominate in the sample.

\cite{Ueda19} reported that the bulk motion induced by line-of-sight gas sloshing in the cool core of A907 is $\sim 800$\,km\,s$^{-1}$, which is smaller than the adiabatic sound speed. Turbulent velocities in cool cores were constrained from X-ray observations with the {\it XMM-Newton} Reflection Grating Spectrometer (RGS). The typical upper limit of the turbulent velocity is $300$\,km\,s$^{-1}$ \citep[e.g.,][]{Sanders10, Sanders11b, Bulbul12, Pinto18, Liu21}, which is also smaller than the sound speed of the ICM. \cite{Ogorzalek17} placed a constraint on the turbulent velocity of the diffuse hot gas in nearby giant elliptical galaxies from a combined analysis of resonance scattering and line broadening using \textit{XMM}/RGS. The inferred turbulent velocity is $\sim 110$\,km\,s$^{-1}$, implying a typical turbulent contribution to nonthermal pressure of $\sim 6$\,\%. Their results obtained for giant elliptical galaxies are comparable to those for galaxy clusters. Alternatively, turbulent motions in the ICM can be constrained by measuring power spectra of fluctuations in the X-ray surface brightness, and the typical estimated turbulent velocities are smaller than $100 - 150$\,km\,s$^{-1}$ within $r\sim 50$\,kpc \citep[e.g.,][]{Zhuravleva14, Zhuravleva18}.
In the future, {\it XRISM} \citep{Tashiro18} and {\it Athena} \citep{Barcons17} will enable gas velocity measurements including both turbulent and bulk motions and will allow us to directly study velocity fields of the ICM.

Detailed multi-probe studies of CLASH galaxy clusters combining X-ray, Sunyaev--Zel'dovich effect (SZE), and gravitational lensing data sets revealed a small fraction of nonthermal pressure support ($P_\mathrm{nth}/P_\mathrm{tot}$) in the cluster cores \citep{Siegel18, Sayers21}. By performing a three-dimensional triaxial analysis of 16 high-mass CLASH clusters, \cite{Sayers21} obtained an upper limit on the ensemble-averaged $P_\mathrm{nth}/P_\mathrm{tot}$ of 9\,\% within $\sim 250$\,kpc.

\cite{Ueda18} studied gas density and pressure perturbations in the cool core of RXJ~1347.5$-$1145  solving the equation of state of perturbations in the hot gas from X-ray and SZE residual images. They found that the perturbation is nearly isobaric, implying that the velocity of gas motions is much smaller than the adiabatic sound speed (i.e., $420^{+310}_{-420}$\,km\,s$^{-1}$). 

Overall, our results are in good agreement with the findings from a range of observations in the literature. These observational facts suggest that a possible mechanism that creates gas perturbations in cool cores must satisfy the following conditions: (1) pressure equilibrium, (2) an isobaric process, (3) low nonthermal pressure (say, $\lesssim 10\%$), and (4) no significant mass dependence. Gas sloshing induced by infalling substructures  satisfies all the above conditions and thus serves as a candidate mechanism for heating gas in cool cores.

\subsection{X-ray brightness contrast versus ICM properties}
\label{sec:thercon}

We measured the X-ray brightness contrast $|\Delta I_{\rm X}| / \langle I_{\rm X} \rangle$ in the cool core of each system, as shown in Figure~\ref{fig:histo} along with the CLASH results of \citet{Ueda20}. The observed distributions for the two samples are similar to each other.

A hypothesis that the cool cores of more massive clusters are less susceptible to disturbance induced by infalling substructures is conceivable. Using hydrodynamic simulations of binary cluster mergers, \cite{ZuHone11b} studied the transition of ICM properties during mergers as a function of merger parameters, such as the initial mass ratio and impact parameter. They found that off-axis mergers with higher mass ratios (e.g., 1:10) causes less impact on the ICM properties. \cite{Valdarnini21} found simular results using their simulations. 

In agreement with the results of \cite{Ueda20}, we find no significant correlation between the X-ray brightness contrast and the ICM properties. A further study is required to robustly examine the correlations using a larger sample. The comparison of the brightness contrast with the ICM properties is given in Appendix~\ref{sec:compcon}.

For low temperature systems with $kT \lesssim 1$~keV where line cooling is dominant, it is expected that the X-ray brightness contrast $|\Delta I_\mathrm{X}| / \langle I_\mathrm{X} \rangle$ is sensitive to metallicity as well as gas density. To investigate this possibility, we focus on three outliers having $|\Delta I_\mathrm{X}| / \langle I_\mathrm{X} \rangle > 1$, namely, NGC~0507, MKW4, and A1644. For NGC~0507 and MKW4 with $kT_\mathrm{pos, neg}<2$\,keV, the metal abundance in the positive excess regions is comparable to that in the negative excess regions (see Table~\ref{tab:fit}). For A1644, the metal abundance in the positive excess region is higher than that of the negative excess region. However, the cool core of A1644 is relatively hot ($kT_\mathrm{pos} \sim 2.7$\,keV and $kT_\mathrm{neg}\sim 4.5$\,keV), so that it is expected that thermal bremsstrahlung dominates over line cooling. We thus find no significant correlation between the X-ray brightness contrast and the metal abundance, as shown in Figure~\ref{fig:kTvsdIx}.

\subsection{Comparison between \Chandra\ observations and numerical simulations}
\label{sec:comp}

Comparisons between observations and numerical simulations are essential to properly interpret data and understand the limitation of observational results. \cite{Ueda20} analyzed synthetic X-ray observations from a high-resolution, hydrodynamic simulation of a binary cluster merger to test the detection algorithm for gas density perturbations. Their analysis, however, was limited to one particular merger configuration with three viewing angles. 

To study thermodynamic characteristics of cool cores for a range of merger configurations, we analyze many sets of synthetic X-ray observations of simulated binary cluster mergers publicly available form the Galaxy Cluster Merger Catalog\footnote{\url{http://gcmc.hub.yt/index.html}} \citep{ZuHone18}. The Galaxy Cluster Merger Catalog offers an extensive suite of synthetic observations based on $N$-body hydrodynamic simulations of binary cluster mergers as well as cosmological simulations of galaxy clusters. Here we focus on their binary merger simulations performed with the FLASH code \citep{ZuHone10, ZuHone11b}, which allow us to study the thermodynamic properties of cluster cores during cluster mergers as a function of merger parameters. 

Two-dimensional maps of cluster observables at redshift $z=0.05$, such as the total density (dark matter and hot gas), gas temperature, and X-ray emissivity, projected along the three cardinal axes ($X,Y,Z$) are made publicly available. Projections were made such that a cluster merger takes place in the $X$--$Y$ plane and the $Z$-axis is perpendicular to the $X$-$Y$ plane. In addition, X-ray event files and images from synthetic 50~ks \Chandra/ACIS-I observations created with {\tt pyXSIM} \citep{ZuHone14} are available. We note that two dimensional projections of gas density and pressure are not included in the public release of the Galaxy Cluster Merger Catalog and are thus not available.

In their simulations, a flat $\Lambda$CDM cosmology with $H_0=71$~km~s$^{-1}$~Mpc$^{-1}$ and $\Omega_\mathrm{m}=0.27$ was assumed for the purposes of calculating distance and redshift-dependent quantities. We note that the pixel size of simulated X-ray images is different from that of \Chandra/ACIS-I but corresponds to the width of the finest simulation cell size (see Table~\ref{tab:sim}). The pixel scale of \Chandra/ACIS-I ($0.492\arcsec$~pixel$^{-1}$) corresponds to $0.48$~kpc at $z=0.05$. In this study, we analyze six different sets of binary merger simulations with different values of the cluster mass, initial mass ratio, and impact parameter, focusing on off-axis minor mergers relevant to the presence of sloshing cool cores. Table~\ref{tab:sim} summarizes the basic information of binary merger simulations used in this work.

%%%%%%%%%%%%%%%%%%%%%%%%%%%%
\begin{table*}[ht]
\begin{center}
\caption{
Summary of binary cluster merger simulations extracted from the Galaxy Cluster Merger Catalog.
}\label{tab:sim}
\begin{tabular}{ccccccl}
\hline\hline
Run	& Primary cluster mass ($M_{200}$)	& Mass ratio	& Impact parameter	& Epoch	& Finest cell size	& References\tablenotemark{a}	\\	
			& (\MO)								&				& (kpc)				& (Gyr)	& (kpc)	&			\\	\hline
Sim1		& $6 \times 10^{14}$				& 1:3			& 500				& 2		& 6.96	& 1			\\
Sim2		& $6 \times 10^{14}$				& 1:3			& 1000				& 2.1\tablenotemark{b}	& 6.96	& 1			\\
Sim3		& $6 \times 10^{14}$				& 1:10			& 500				& 2		& 6.96	& 1			\\
Sim4		& $6 \times 10^{14}$				& 1:10			& 1000				& 2		& 6.96	& 1			\\
Sim5		& $1 \times 10^{15}$				& 1:20			& 200				& 2		& 4.88	& 2, 3		\\
Sim6		& $1 \times 10^{15}$				& 1:20			& 1000				& 2		& 4.88	& 2, 3		\\
\hline
\end{tabular}
\end{center}
\tablenotetext{a}{\small
1 -- \cite{ZuHone11b}, 2 -- \cite{ZuHone10}, 3 -- \cite{ZuHone16}
}
\tablenotetext{b}{\small
Since the simulation data at the epoch of 2\,Gyr are not available, we adopted those at 2.1\,Gyr instead. 
}
\end{table*}
%%%%%%%%%%%%%%%%%%%%%%%%%%%%

We have analyzed synthetic \Chandra\ data of binary cluster mergers to obtain their residual images by subtracting the global brightness profile using the same method as presented in Section~\ref{subsec:ximg}. To investigate systematic effects associated with the choice of the center for ellipse modeling, we adopt three different definitions of the projected cluster center, namely the peak of the total mass distribution, the peak of the X-ray emissivity map, and the peak of the simulated X-ray image. For each projection, Table~\ref{tab:po} lists the absolute mean offset and standard deviation of the positional offset between the projected mass and X-ray peaks. Here we consider two X-ray peak definitions, namely the emissivity peak and the peak in the synthetic X-ray image. The statistics for each projection is calculated combining all merger configurations. 
We find a typical positional offset of $\sigma \sim 15$\,kpc, which is $\sim 3$ times larger than the finest cell size of the simulations.

%%%%%%%%%%%%%%%%%%%%%%%%%%%%
\begin{table}
\begin{center}
\caption{Statistics of the positional offset between the projected total mass and X-ray peak positions derived from binary cluster merger simulations.
}\label{tab:po}
\begin{tabular}{lccc}
\hline\hline	
Center	& Projection	& Mean\tablenotemark{a} 	& Std\tablenotemark{b}		\\ 
			&		& (kpc)	& (kpc)		\\	\hline
X-ray emissivity map\tablenotemark{c}	& X				& 12.23			& 11.18		\\
			& Y				& 14.61			& 15.34		\\
			& Z				& 18.59			& 18.12		\\
Synthetic {\it Chandra} map\tablenotemark{d}	& X				& 13.52			& 11.49		\\
			& Y				& 14.31			& 15.27		\\
			& Z				& 18.75			& 17.07		\\
\hline
\end{tabular}
\end{center}
\tablenotetext{a}{Absolute mean offset, $\langle|
\mbox{\boldmath{$d$}}|\rangle$.}
\tablenotetext{b}{Standard deviation, $\sigma(\mbox{\boldmath{$d$}})$.}
\tablenotetext{c}{Noise-free X-ray emissivity maps.}
\tablenotetext{d}{Synthetic \Chandra\ X-ray maps.}
\end{table}
%%%%%%%%%%%%%%%%%%%%%%%%%%%%

We find a pair of positive and negative excess regions in the X-ray residual images for all configurations and viewing angles analyzed. We determine the ICM temperature in each perturbed region and measure the X-ray brightness contrast of density perturbations $|\Delta I_{\rm X}| / \langle I_{\rm X} \rangle$ for each projection angle and center definition. These results are summarized in Tables~\ref{tab:simkT} and \ref{tab:simIx}.

We find that the temperature ratio $T_\mathrm{neg}/T_\mathrm{pos}$ between the negative and positive excess regions is systematically higher than unity, with a clear offset between both regions. The mean temperature ratios are obtained as $1.37 \pm 0.16$, $1.30 \pm 0.15$, or $1.33 \pm 0.14$, for ellipse modeling with the total mass peak, emissivity peak, and X-ray image peak, respectively. The results are comparable to the observed ratio of $T_\mathrm{neg}/T_\mathrm{pos} = 1.20^{+0.04}_{-0.03}$. We thus find no significant dependence of the chosen center from synthetic observations of the off-axis merger simulations. 

Figure~\ref{fig:kTmvsdIx} shows a comparison of the X-ray brightness contrast with the ICM temperature, which is similar to our observational results (Table~\ref{tab:dIx} and Figure~\ref{fig:kTvsdIx}). Although the simulations analyzed only cover a limited range of parameter space of mergers,  the comparison shows that gas sloshing inducted by infalling substructures can recover characteristic gas density perturbations observed in strong cool-core systems.

%%%%%%%%%%%%%%%%%%%%%%%%%%%%
\begin{table*}
\begin{center}
\caption{Temperatures extracted from perturbed regions in X-ray residual images obtained using ellipse modeling with respect to the projected mass peak, the X-ray emissivity peak, or the peak of the synthetic {\it Chandra} X-ray image.
}\label{tab:simkT}
\begin{tabular}{lcccccccccc}
\hline\hline
				& 	& \multicolumn{3}{c}{Total mass peak}	& \multicolumn{3}{c}{Emissivity peak}	& \multicolumn{3}{c}{X-ray image peak}	\\	
Run		& Proj.	& Positive	& Negative	& Ratio	& Positive	& Negative	& Ratio	& Positive	& Negative	& Ratio	\\
				&				& (keV)		& (keV)		& 		& (keV)		& (keV)		&		& (keV)		& (keV)		&		\\ \hline
Sim1 & X & $3.98 \pm 0.66$ & $6.26 \pm 0.56$ & $1.57 \pm 0.30$ & $4.42 \pm 0.23$ & $5.83 \pm 0.53$ & $1.32 \pm 0.14$ & $4.38 \pm 0.26$ & $5.78 \pm 0.60$ & $1.32 \pm 0.16$ \\
	& Y & $4.23 \pm 0.71$ & $5.94 \pm 0.81$ & $1.40 \pm 0.30$ & $5.01 \pm 0.53$ & $4.94 \pm 0.38$ & $0.99 \pm 0.13$ & $4.10 \pm 0.21$ & $5.17 \pm 0.18$ & $1.26 \pm 0.08$ \\
	& Z & $3.95 \pm 0.84$ & $6.78 \pm 0.61$ & $1.72 \pm 0.40$ & $4.37 \pm 0.32$ & $6.14 \pm 0.51$ & $1.41 \pm 0.16$ & $4.31 \pm 0.31$ & $6.10 \pm 0.51$ & $1.42 \pm 0.16$ \\
Sim2 & X & $4.51 \pm 0.67$ & $6.42 \pm 0.68$ & $1.42 \pm 0.26$ & $4.50 \pm 0.56$ & $5.70 \pm 0.54$ & $1.27 \pm 0.20$ & $3.71 \pm 0.31$ & $5.64 \pm 0.53$ & $1.52 \pm 0.19$ \\
	& Y & $3.65 \pm 0.29$ & $5.05 \pm 0.78$ & $1.38 \pm 0.24$ & $4.26 \pm 0.53$ & $4.57 \pm 0.52$ & $1.07 \pm 0.18$ & $4.26 \pm 0.53$ & $4.57 \pm 0.52$ & $1.07 \pm 0.18$ \\
	& Z & $3.90 \pm 0.53$ & $6.68 \pm 0.83$ & $1.71 \pm 0.32$ & $4.37 \pm 0.57$ & $5.94 \pm 0.67$ & $1.36 \pm 0.23$ & $4.05 \pm 0.39$ & $5.86 \pm 0.70$ & $1.45 \pm 0.22$ \\
Sim3 & X & $4.50 \pm 0.56$ & $5.70 \pm 0.54$ & $1.27 \pm 0.20$ & $3.96 \pm 0.23$ & $6.38 \pm 0.73$ & $1.61 \pm 0.21$ & $3.96 \pm 0.23$ & $6.38 \pm 0.73$ & $1.61 \pm 0.21$ \\
	& Y & $4.29 \pm 0.51$ & $4.95 \pm 0.71$ & $1.15 \pm 0.21$ & $4.97 \pm 0.83$ & $5.09 \pm 0.55$ & $1.02 \pm 0.20$ & $4.97 \pm 0.83$ & $5.09 \pm 0.55$ & $1.02 \pm 0.20$ \\
	& Z & $4.52 \pm 0.62$ & $5.93 \pm 0.74$ & $1.31 \pm 0.24$ & $3.80 \pm 0.46$ & $6.78 \pm 0.65$ & $1.78 \pm 0.28$ & $3.80 \pm 0.42$ & $6.86 \pm 0.60$ & $1.81 \pm 0.25$ \\
Sim4 & X & $4.33 \pm 0.31$ & $5.76 \pm 0.52$ & $1.33 \pm 0.15$ & $4.16 \pm 0.67$ & $6.24 \pm 0.60$ & $1.50 \pm 0.28$ & $4.24 \pm 0.65$ & $6.21 \pm 0.64$ & $1.46 \pm 0.27$ \\
	& Y & $4.30 \pm 0.40$ & $4.84 \pm 0.43$ & $1.13 \pm 0.14$ & $4.23 \pm 0.71$ & $5.94 \pm 0.81$ & $1.40 \pm 0.30$ & $4.23 \pm 0.71$ & $5.94 \pm 0.81$ & $1.40 \pm 0.30$ \\
	& Z & $4.44 \pm 0.42$ & $6.00 \pm 0.49$ & $1.35 \pm 0.17$ & $4.05 \pm 0.88$ & $6.73 \pm 0.63$ & $1.66 \pm 0.39$ & $4.05 \pm 0.88$ & $6.73 \pm 0.63$ & $1.66 \pm 0.39$ \\
Sim5 & X & $7.22 \pm 0.52$ & $8.82 \pm 0.26$ & $1.22 \pm 0.10$ & $7.15 \pm 0.36$ & $8.86 \pm 0.77$ & $1.24 \pm 0.12$ & $7.19 \pm 0.29$ & $8.85 \pm 0.75$ & $1.23 \pm 0.12$ \\
	& Y & $7.50 \pm 0.55$ & $8.36 \pm 0.31$ & $1.11 \pm 0.09$ & $6.74 \pm 0.60$ & $7.52 \pm 0.63$ & $1.12 \pm 0.14$ & $6.74 \pm 0.60$ & $7.52 \pm 0.63$ & $1.12 \pm 0.14$ \\
	& Z & $7.36 \pm 0.63$ & $9.46 \pm 0.30$ & $1.29 \pm 0.12$ & $7.11 \pm 0.46$ & $9.03 \pm 0.66$ & $1.27 \pm 0.12$ & $7.03 \pm 0.53$ & $9.01 \pm 0.63$ & $1.28 \pm 0.13$ \\
Sim6 & X & $5.49 \pm 0.52$ & $8.71 \pm 0.72$ & $1.59 \pm 0.20$ & $7.45 \pm 0.64$ & $8.55 \pm 0.42$ & $1.15 \pm 0.11$ & $7.45 \pm 0.64$ & $8.71 \pm 0.32$ & $1.17 \pm 0.11$ \\
	& Y & $7.29 \pm 0.82$ & $7.51 \pm 0.64$ & $1.03 \pm 0.15$ & $8.42 \pm 0.48$ & $8.63 \pm 0.31$ & $1.02 \pm 0.07$ & $8.44 \pm 0.45$ & $8.65 \pm 0.43$ & $1.02 \pm 0.07$ \\
	& Z & $5.43 \pm 0.53$ & $8.90 \pm 0.62$ & $1.64 \pm 0.20$ & $8.20 \pm 0.85$ & $9.29 \pm 0.46$ & $1.13 \pm 0.13$ & $8.28 \pm 0.80$ & $9.24 \pm 0.48$ & $1.12 \pm 0.12$ \\
\hline
\end{tabular}
\end{center}
\end{table*}
%%%%%%%%%%%%%%%%%%%%%%%%%%%%

%%%%%%%%%%%%%%%%%%%%%%%%%%%%
\begin{table}
\begin{center}
\caption{
Statistics of the X-ray brightness contrast in perturbed cluster cores extracted from synthetic \Chandra\ X-ray images of binary cluster mergers.
}\label{tab:simIx}
\begin{tabular}{lccc}
\hline\hline	
Center\tablenotemark{a}		& Projection	& \multicolumn{2}{c}{$|\Delta I_{\rm X}| / \langle I_{\rm X} \rangle$}	\\ 
			&				& Mean	& Std		\\	\hline
Total mass peak & X				& 0.277 & 0.078		\\
			& Y				& 0.179 & 0.062		\\
			& Z				& 0.331 & 0.098		\\
Emissivity peak& X				& 0.349 & 0.065		\\
			& Y				& 0.230 & 0.080		\\
			& Z				& 0.415 & 0.127		\\
X-ray image peak & Z				& 0.426 & 0.115		\\
\hline
\end{tabular}
\end{center}
\tablenotetext{a}{Definition of the cluster center chosen for ellipse modeling of the X-ray brightness distribution.}
\end{table}
%%%%%%%%%%%%%%%%%%%%%%%%%%%%

\subsection{ICM metal abundance}
\label{sec:abund}

As shown in the top right panel of Figure~\ref{fig:para}, we find that the ICM abundance in cool cores has a $Z_\mathrm{neg}$--$Z_\mathrm{pos}$ relation that is steeper than the one-to-one relation. Overall, we see a trend that $Z_\mathrm{neg}/Z_\mathrm{pos} \gtrsim 1$ at $Z_\mathrm{pos} \gtrsim 0.8Z_\odot$ and $Z_\mathrm{neg}/Z_\mathrm{pos} \lesssim 1$ at $Z_\mathrm{pos} \lesssim 0.8Z_\odot$. The trend in the high abundance regime is consistent with what is expected by gas sloshing, whereas the trend in the low abundance regime is opposite. As discussed in Section~\ref{sec:thermo}, the correlation of the ICM abundance between the positive and negative excess regions is still under debate.

For low-temperature systems where the Fe-L line complex is the dominant component in the iron line emissions in X-ray spectra, there is a possibility that the measurement of the ICM metal abundance is affected by plasma code uncertainties \citep[e.g.,][]{Mernier20b}. To examine this possibility, we closely follow the procedure outlined in \cite{Ghizzardi21}.  Here we reanalyze X-ray spectra of the lowest temperature system in our sample, namely, NGC~4636, after masking the rest-frame $0.9-1.3$\,keV energy band associated with the Fe-L line complex. We find that the metal abundances in the positive and negative excess regions are $0.350^{+0.015}_{-0.014} Z_\odot$ and $0.686^{+0.055}_{-0.049} Z_\odot$, respectively. These are both consistent with the corresponding measurements without masking (see Table~\ref{tab:fit}). Hence, we conclude that the plasma code uncertainties are not expected to significantly bias metal abundance measurements for our sample.

\cite{Mantz17} analyzed a sample of 245 massive galaxy clusters with $kT > 5$\,keV, finding that cooler cores are more metal enriched. On the other hand, they found that the ICM metal abundance outside cool cores weakly correlates with the core-excised mean temperature, implying that more massive clusters tend to have a higher abundance in the ambient ICM. 
Similarly, \cite{Rasmussen07, Rasmussen09} studied 15 nearby galaxy groups and found that the ICM abundance correlates with the core-excised mean ICM temperature. By analyzing a sample of 207 nearby clusters and groups, \cite{Lovisari19} found that the ICM abundance in the core region is weakly anticorrelate with the temperature measured in the same core region.

To examine the ICM metal abundance in the central regions over a wide mass range, \cite{Mernier16, Mernier17} studied the Chemical Enrichment RGS Sample \citep[CHEERS;][]{de_Plaa17} containing 44 nearby galaxy clusters, groups, and massive elliptical galaxies. They classified the CHEERS sample into two subsamples, namely the "cool" subsample with $kT < 1.7$\,keV mostly corresponding to galaxy groups and massive elliptical galaxies (21 objects) and the "hot" subsample with $kT > 1.7$\,keV (23 objects). They found that the ICM abundance of the central region in the hot sample is higher than that in the cool sample. 

By compiling previous studies, it is expected that (1) the metal abundance of the ambient ICM increases with the mean temperature (i.e., a proxy for the total mass), (2) the cooler the core, the higher the metal abundance, and (3) Galaxy groups and clusters may have different trends of the metal abundance against the temperature. Moreover, a large intrinsic scatter was found for the ICM abundance in cool cores \citep[e.g.,][]{Mernier17}. 

\begin{figure}
 \begin{center}
  \includegraphics[scale=0.33,clip]{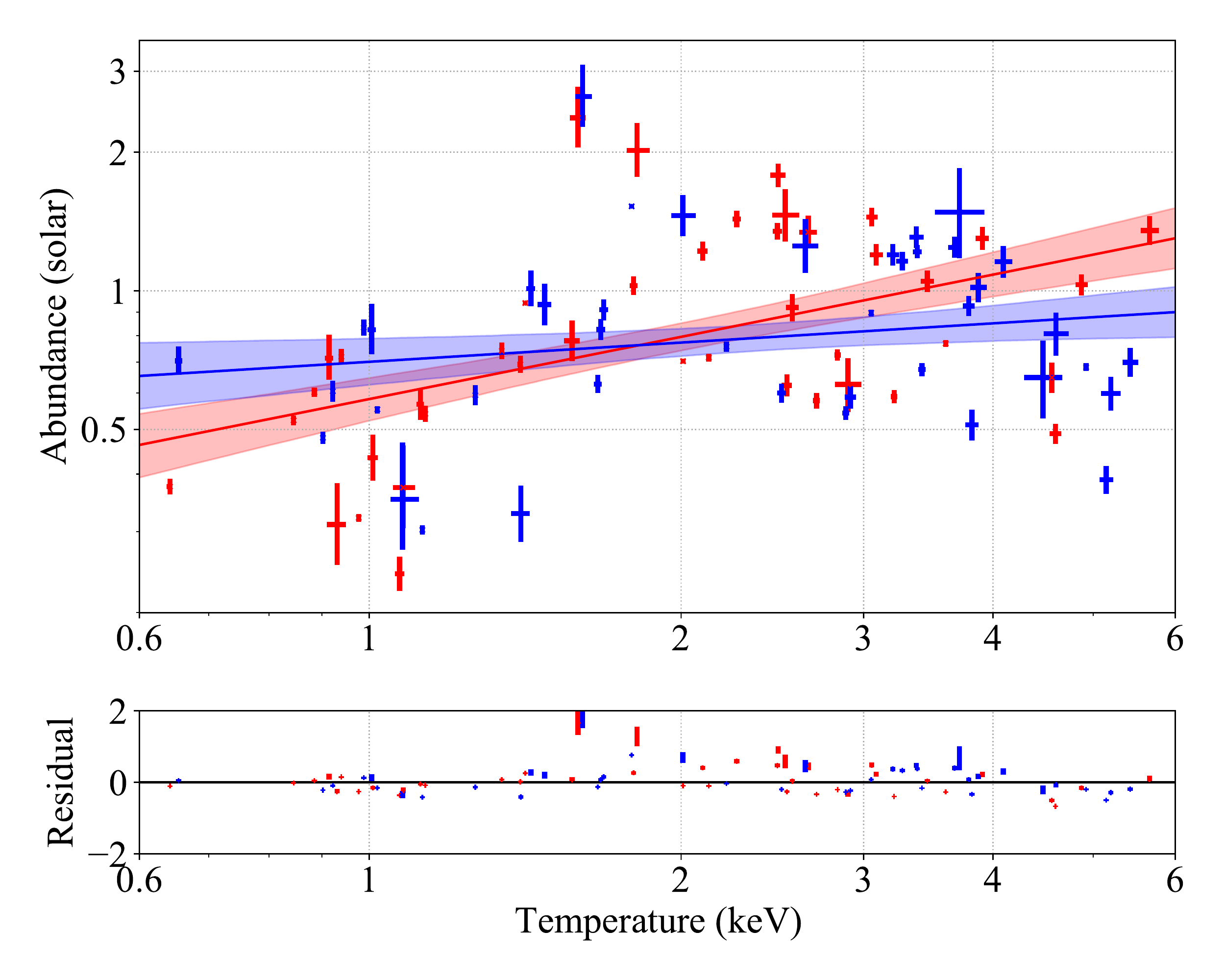}
 \end{center}
\caption{ICM metal abundance versus ICM temperature measured in the positive (red) and negative (blue) excess regions. The red and blue error bars show our measurements extracted from the positive and negative excess regions, respectively. The red and blue shaded regions show the 68\% confidence interval around the best-fit relation for the positive and negative excess regions, respectively.
}
\label{fig:kTvsAbund}
\end{figure}

Figure~\ref{fig:kTvsAbund} shows a comparison between the ICM metal abundance and the ICM temperature in cool cores for the respective excess regions. For both positive and negative excess regions, we find a positive correlation between the ICM abundance and temperature. The abundance--temperature relation ($Z_\mathrm{pos}$--$T_\mathrm{pos}$ or $Z_\mathrm{neg}$--$T_\mathrm{neg}$) is steeper for the positive excess region, and the ICM abundance in the negative excess region only mildly increases with temperature. Since the negative excess regions are expected to be dominated by the ambient ICM, the observed trend appears to be consistent with the previous studies of galaxy clusters. Besides, Figure~\ref{fig:kTvsAbund} shows that, on average, the ICM abundance of the positive excess region lies increasingly below (above) that of the negative excess region at temperatures below (above) $\sim 2$~keV. These different temperature trends in both excess regions are expected to contribute to the large intrinsic scatter of the ICM abundance in cool cores.

\begin{figure}
 \begin{center}
  \includegraphics[scale=0.33,clip]{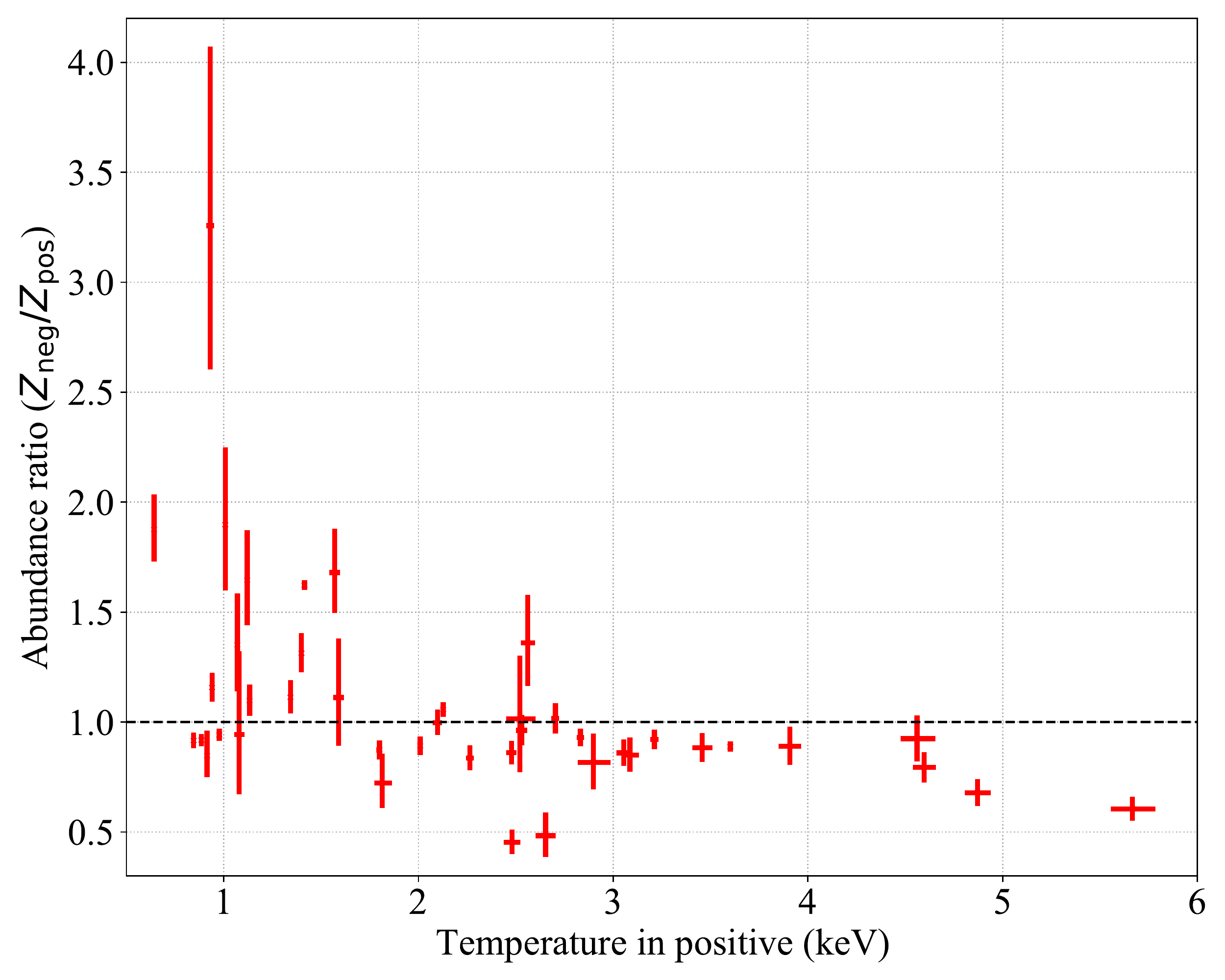}
 \end{center}
\caption{ICM metal abundance ratio, $Z_{\mathrm{neg}}/Z_{\mathrm{pos}}$, as a function of ICM temperature in the positive excess region. The red error bars show our measurements extracted from the perturbed regions in cool cores. The horizontal dashed line corresponds to unity.
}
\label{fig:kTvsAbundRatio}
\end{figure}

To examine the trend of ICM metal abundance with temperature, we plot in Figure~\ref{fig:kTvsAbundRatio} the ratio $Z_\mathrm{neg}/Z_\mathrm{pos}$ of the ICM abundance as a function of temperature $kT_\mathrm{pos}$ in the positive excess region.  In the regime where $kT_\mathrm{pos} \gtrsim  2$~keV, the abundance ratios tend to be lower than unity, while the scatter appears to increase at temperatures below $kT_\mathrm{pos} \sim 2$\,keV. This is consistent with the trend in Figure~\ref{fig:kTvsAbund} at $\gtrsim 2$~keV and may explain the best-fit $Z_\mathrm{neg}$--$Z_\mathrm{pos}$ slope shallower than unity.

\begin{figure}
 \begin{center}
  \includegraphics[scale=0.33,clip]{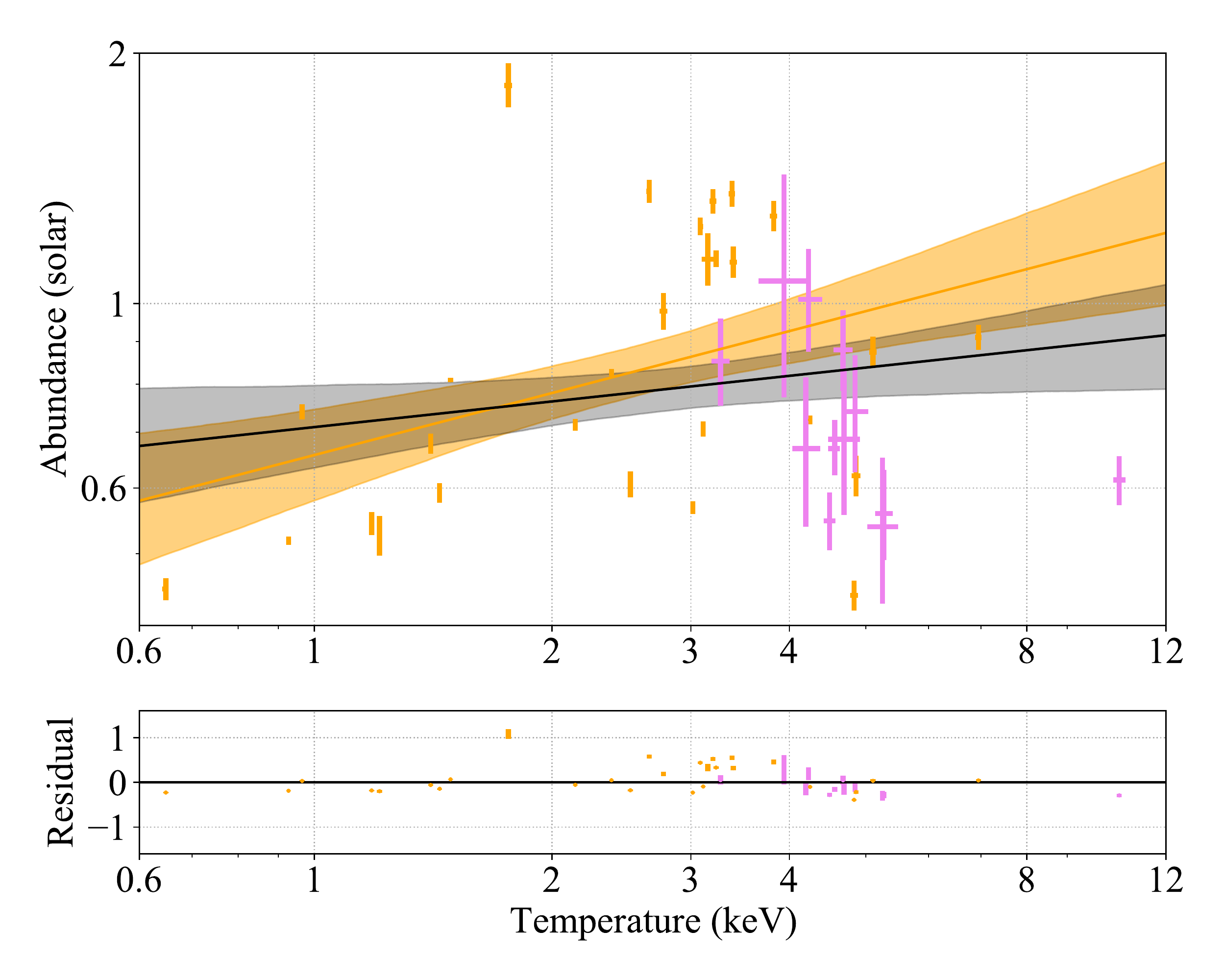}
 \end{center}
\caption{Same as Figure~\ref{fig:kTvsAbund} but for the mean values extracted from the whole perturbed region in each cool-core system. The orange and magenta error bars represent the measurements for the sample in this work (HIFLUGCS) and the CLASH sample \citep{Ueda20}, respectively. The orange shaded region indicates the 68\% confidence interval around the best-fit relation (solid line) for the present sample. Similarly, the gray shaded region show the results for the composite sample. The best-fit residuals for the composite sample are shown in the lower panel.
}
\label{fig:kTvsAbund2}
\end{figure}

In Figure~\ref{fig:kTvsAbund2}, we compare the mean ICM temperature and metal abundance extracted from the cool-core regions for our sample. In this comparison, we also show the results for the CLASH cool-core sample of \citet{Ueda20}. In the figure, we show our regression results of the $Z$--$T$ relations with and without the CLASH sample. We find a clear positive correlation for the HIFLUGCS sample. When the CLASH sample is included, the composite sample exhibits a positive but much weaker trend with temperature. In fact, we find a negative temperature slope for the CLASH-only $Z$--$T$ relation. Hence, as found by previous studies, the ICM abundance appears to have a complex dependence on the temperature in cool cores, and there is a likely transition between group--cluster scales (say, at $kT=2$--$3$~keV). A large homogeneous sample spanning the full range of group--cluster scales is needed to robustly address this problem.

%%%%%%%%%%%%%%%%%%%%%%%%%%%%
%%%%%%%%%%%%%%%%%%%%%%%%%%%%

%%%%%%%%%%%%%%%%%%%%%%%%%%%%
%%%%%%%%%%%%%%%%%%%%%%%%%%%%

%%%%%%%%%%%%%%%%%%%%%%%%%%%%
%%%%%%%%%%%%%%%%%%%%%%%%%%%%

\section{Summary and Conclusions}
\label{sec:conclusions}

In this paper, we have conducted a systematic study of gas density perturbations detected in the central regions of 28 strong cool-core systems selected from the HIghest X-ray FLUx Galaxy Cluster Sample (HIFLUGCS)  using archival X-ray data from  the {\it Chandra X-ray Observatory}. The goal of this study was to investigate possible origins of the characteristic gas perturbations and to examine the role of gas sloshing as a possible heating source of cool cores.
To this end, we have characterized the thermodynamic properties of hot gas in the perturbed regions and established their correlations between positive and negative excess regions. The main conclusions of this paper are summarized as follows:

\begin{enumerate}

\item All individual systems in our sample selected from HIFLUGCS exhibit at least a pair of positive and negative excess regions in their X-ray residual image after subtracting the global brightness profile. Since our sample contains not only galaxy clusters but also galaxy groups and massive elliptical galaxies (central dominant galaxies in groups and clusters), the characteristic dipolar perturbations appear to be a universal feature in strong cool cores.

\item We performed a regression analysis of thermodynamic properties of the ICM between both excess regions, finding that the best-fit relations for temperature and entropy show a clear offset from the one-to-one relation: $T_\mathrm{neg}/T_\mathrm{pos} = 1.20^{+0.04}_{-0.03}$ and $K_\mathrm{neg}/K_\mathrm{pos}= 1.43 \pm 0.07$. The best-fit relation for pressure is found to be consistent with the one-to-one relation, $P_\mathrm{neg} = P_\mathrm{pos}$.

\item To improve statistics and cover a wider mass range, we combined our sample with a high-mass cool-core sample from CLASH \citep{Ueda20}. The composite sample spans a mass range of $M_{500}\in [10^{13}$, $10^{15}] M_\odot$.
We found that the best-fit $P_\mathrm{neg}$--$P_\mathrm{pos}$ relation for the composite sample is remarkably consistent with the one-to-one relation with an intrinsic scatter of $14\pm 2\,\%$. indicating that the perturbed regions are in pressure equilibrium. Our finding supports the hypothesis that the observed gas density perturbations in cool cores are generated by subsonic gas sloshing.

\item For each individual system in our sample, we measured the X-ray brightness contrast of fluctuations in the positive and negative excess regions. The observed distribution of the X-ray brightness contrast is similar to that of the CLASH cool-core sample. We found three outliers, namely NGC~0507, MKW4, and A1644, whose brightness contrast exceeds unity. NGC~0507 and MKW4 are low mass, galaxy--group scale objects, whereas A1644 is a post merger exhibiting a clear spiral-like X-ray morphology in the central region.

\item To understand the observed characteristics in the cool cores, we analyzed many sets of synthetic observations of perturbed cluster cores created from binary cluster merger simulations \citep{ZuHone18}. In all cases, we detected a pair of positive and negative excess regions in synthetic \Chandra\ X-ray images. The temperature ratios between both regions, $T_\mathrm{neg}/T_\mathrm{pos}\sim 1.3$, are in agreement with our measurements, and the distribution of X-ray brightness contrast is consistent with our results. The observed ICM characteristics in strong cool cores can thus be explained by subsonic gas sloshing caused by infalling substructures.

\end{enumerate}

\begin{acknowledgments}
We are grateful to the anonymous referee for helpful suggestions and comments.
This work made use of data from the Galaxy Cluster Merger Catalog (\url{http://gcmc.hub.yt}).
F.L.N. acknowledges the ASIAA Summer School 2019 for giving an opportunity to participate research in ASIAA and for their hospitality. 
The scientific results reported in this article are based on data obtained from the Chandra Data Archive.
This work is supported in part by the Ministry of Science and Technology of Taiwan (grants MOST 106-2628-M-001-003-MY3 and 109-2112-M-001-018-MY3) and by Academia Sinica (grant AS-IA-107-M01). 
Y.I. is supported by the Grants-in-Aid for Scientific Research by the Japan Society for the Promotion of Science (JSPS) with KAKENHI grant Nos JP18H05458, JP20K14524, and JP20K20527.
T.K. acknowledges support from Grant-in-Aid for Scientific Research by JSPS KAKENHI grant number JP18K03704.
\end{acknowledgments}

%\clearpage
\appendix
\section{Comparisons of the X-ray brightness contrast with ICM properties}
\label{sec:compcon}

\begin{figure*}
 \begin{center}
  \includegraphics[scale=0.4,clip]{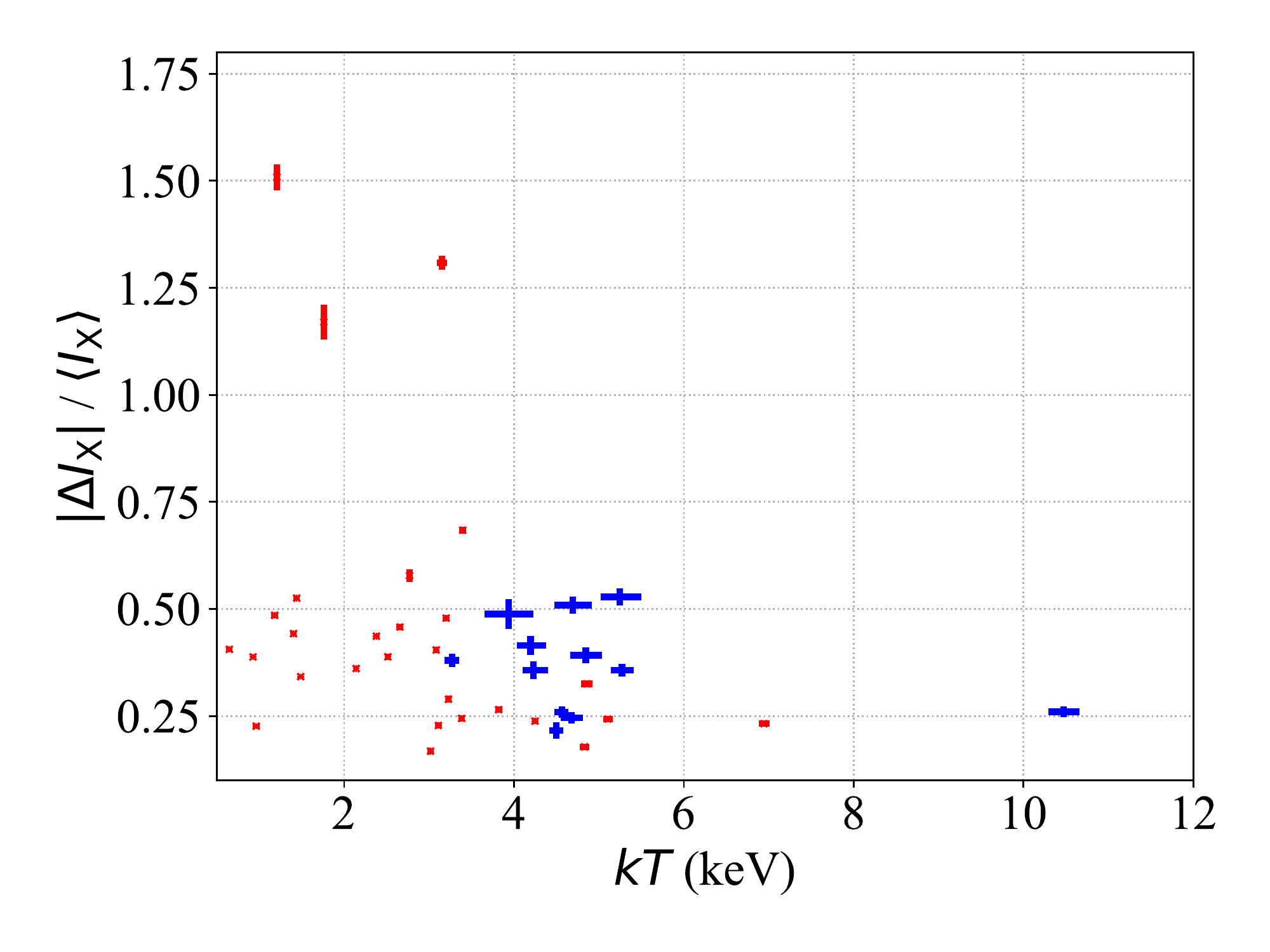}
  \includegraphics[scale=0.4,clip]{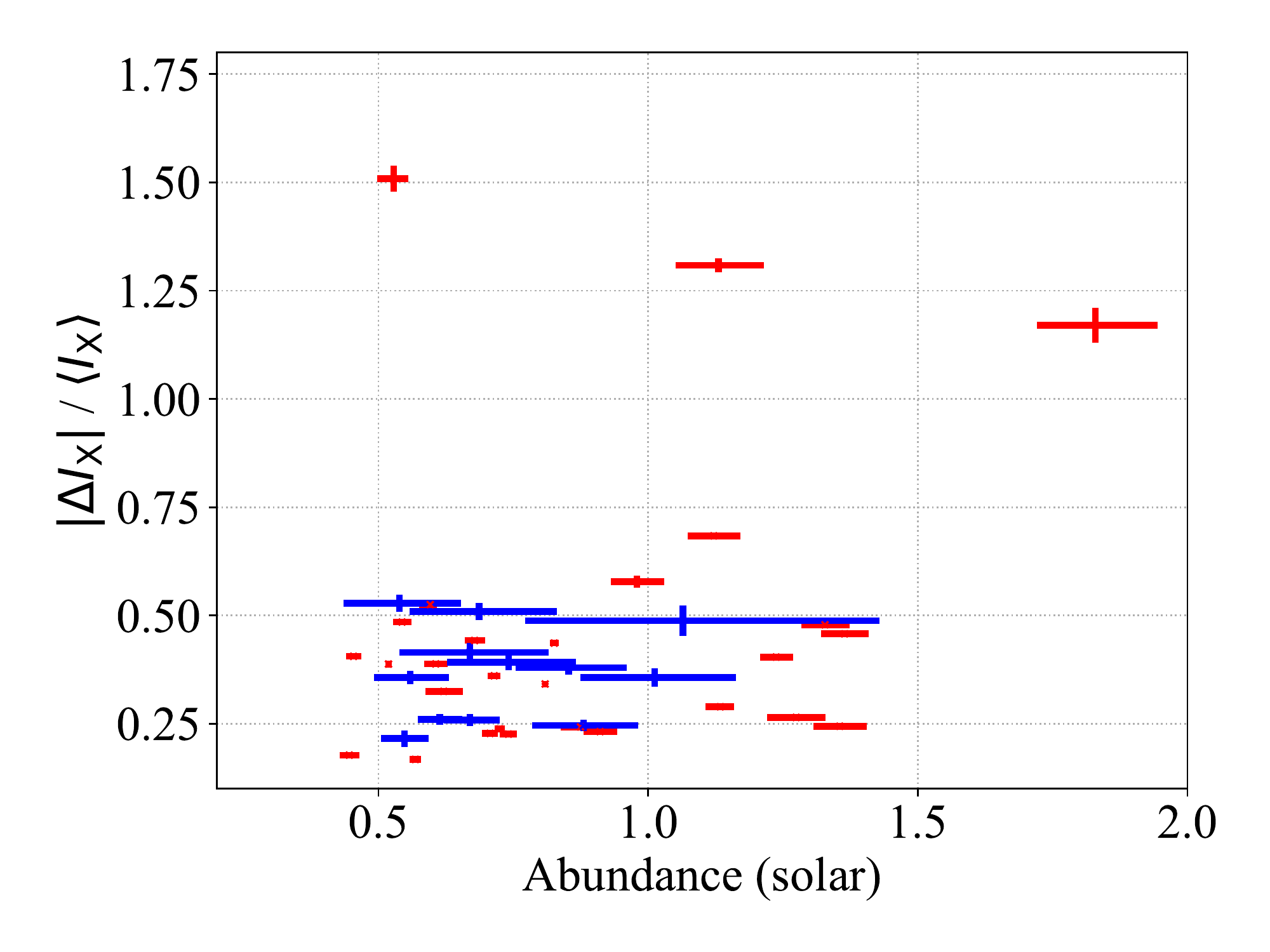}
  \includegraphics[scale=0.4,clip]{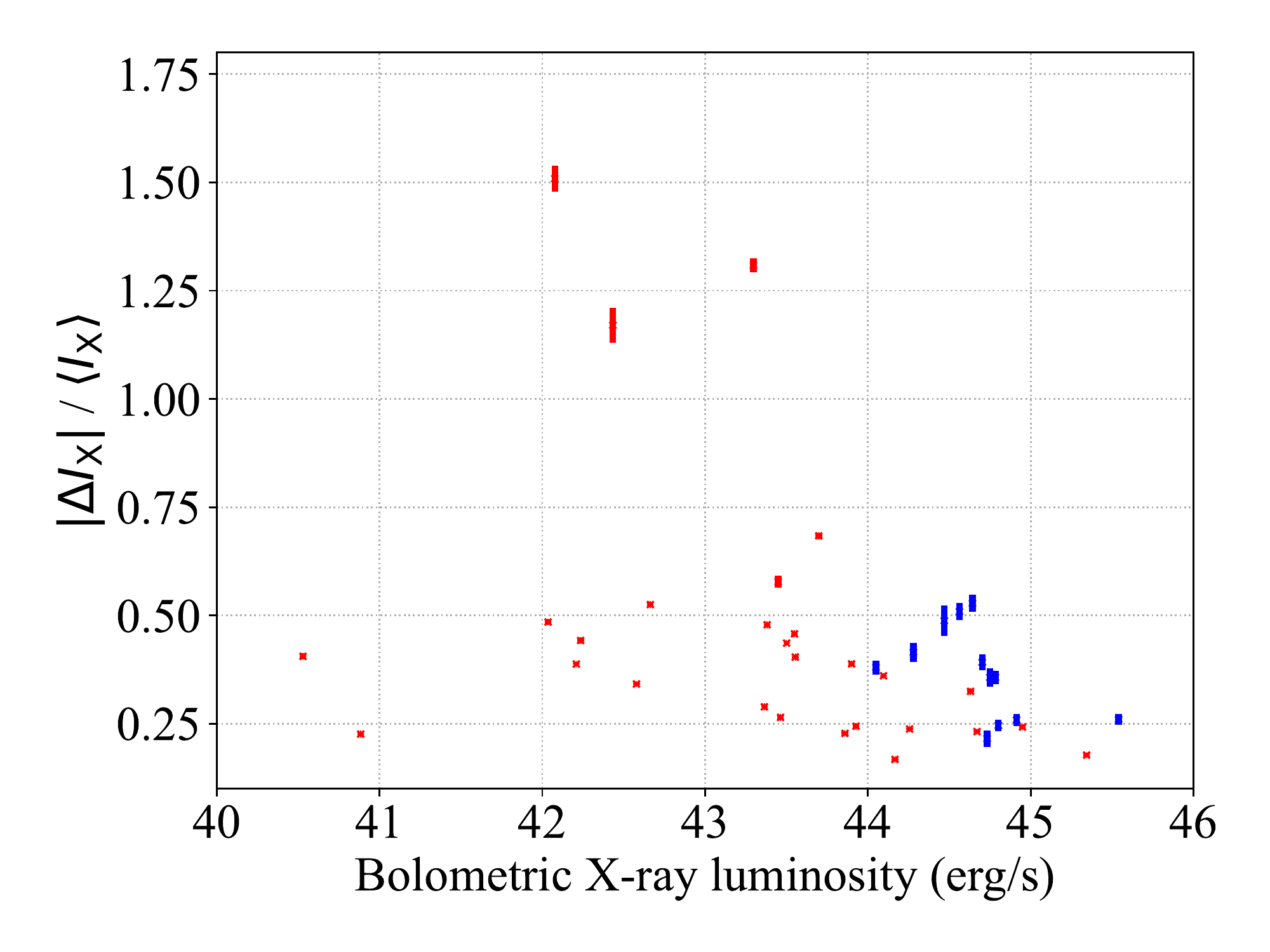}
 \end{center}
\caption{X-ray brightness contrast $|\Delta I_{\rm X}| / \langle I_{\rm X} \rangle$ shown as a function of the ICM temperature (top left), metal abundance (top right), or logarithm of the bolometric X-ray luminosity in the $0.001 - 100$\,keV band (bottom). The measurements for our sample are shown with red error bars, while the measurements for the CLASH cool-core sample \citep{Ueda20} are shown with blue error bars.
}
\label{fig:kTvsdIx}
\end{figure*}

\begin{figure}
 \begin{center}
  \includegraphics[width=6.5cm]{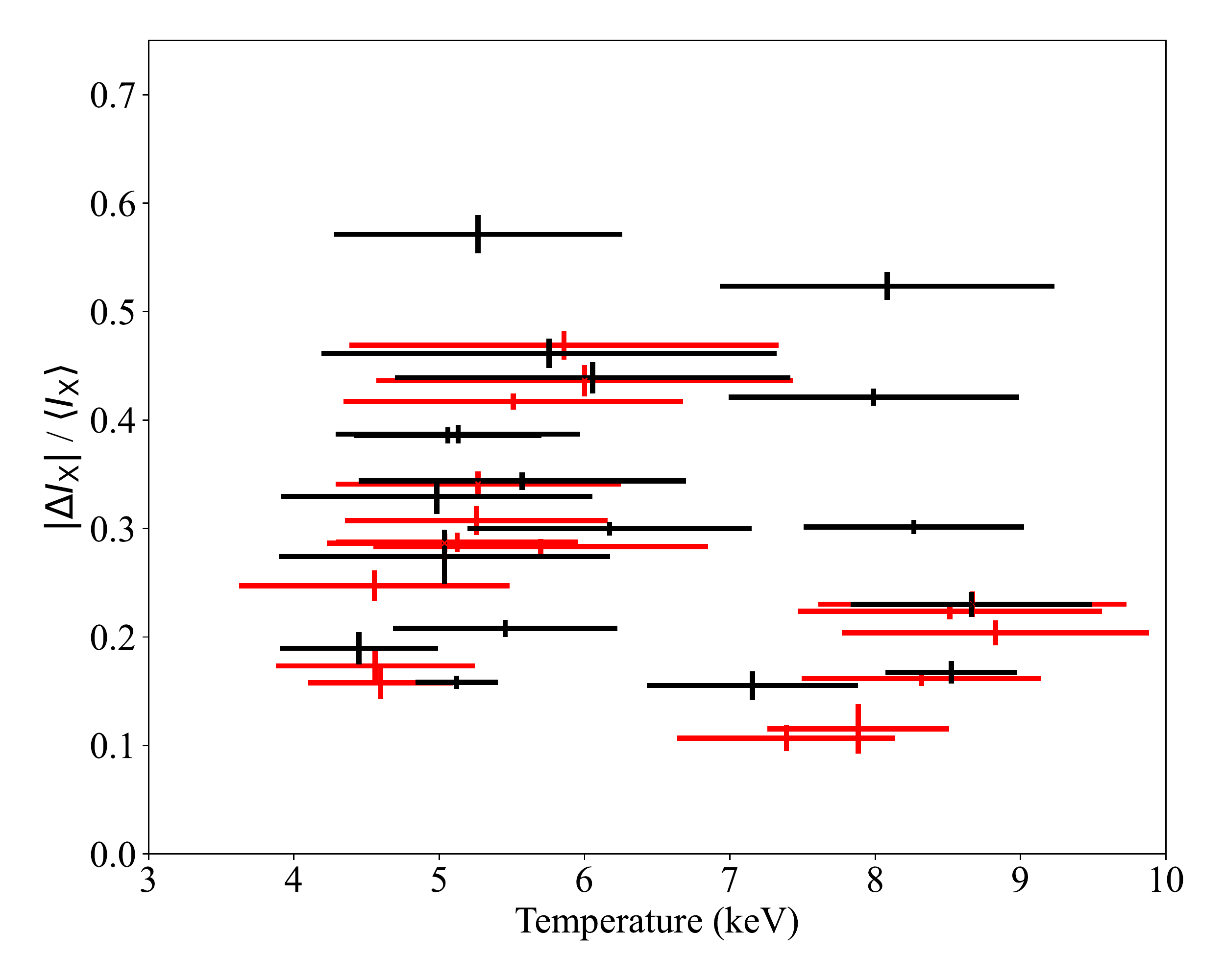}
 \end{center}
\caption{X-ray brightness contrast $|\Delta I_{\rm X}| / \langle I_{\rm X} \rangle$ as a function of temperature from synthetic \Chandra\ observations. The red and black error bars show the results obtained using X-ray residual images extracted with respect to the total mass peak and the X-ray image peak, respectively.
}
\label{fig:kTmvsdIx}
\end{figure}

Here we compare for our sample the X-ray brightness contrast with thermodynamic properties of the ICM, namely the mean temperature, metal abundance, and X-ray luminosity in the $0.001 - 100$\,keV band corresponding to the bolometric X-ray luminosity (Figure~\ref{fig:kTvsdIx}). The results for the CLASH cool-core sample of \cite{Ueda20} are also shown. A comparison between the brightness contrast and temperature from binary merger simulations is also shown in Figure~\ref{fig:kTmvsdIx}. Overall, no significant correlation is found between the X-ray brightness contrast and the thermodynamic ICM properties.

%%%%%%%%%%%%%%%%%%%%%%%%%%%%
%%%%%%%%%%%%%%%%%%%%%%%%%%%%

%%%%%%%%%%%%%%%%%%%%%%%%%%%%
%%%%%%%%%%%%%%%%%%%%%%%%%%%%

% See the manual for the detail.
%%%

\bibliographystyle{aasjournal}
%\bibliography{00_BibTeX_library}
%\bibliography{draft.bbl}
\bibliography{draft}

\end{document}

%% file: table/table_kTa.tex
\begin{deluxetable*}{lccccc}
\tablecaption{
Best-fit parameters of the X-ray spectral analysis performed in each of the perturbed regions identified in X-ray residual images. Projected temperatures and metal abundances extracted from individual perturbed regions are listed in the table.
\label{tab:fit}
}
\tablehead{
\colhead{Cluster} &
\colhead{ID\tablenotemark{a}}	&
\multicolumn{2}{c}{Positive excess region\tablenotemark{b}}	& 
\multicolumn{2}{c}{Negative excess region\tablenotemark{b}}	\\
\colhead{}		&
\colhead{}		&
\colhead{$kT$	(keV)}			&
\colhead{Abundance (\ZO)}	&
\colhead{$kT$ (keV)}	&
\colhead{Abundance (\ZO)}
}
\startdata
             A85 &    1 & $2.521_{-0.071}^{+0.080}$ & $1.460_{-0.179}^{+0.205}$ & $3.714_{-0.196}^{+0.213}$ & $1.482_{-0.303}^{+0.366}$ \\
             	 &    2 & $3.907_{-0.058}^{+0.058}$ & $1.300_{-0.073}^{+0.076}$ & $4.093_{-0.077}^{+0.085}$ & $1.157_{-0.089}^{+0.094}$ \\
            A133 &    1 & $2.264_{-0.020}^{+0.020}$ & $1.432_{-0.058}^{+0.061}$ & $3.203_{-0.044}^{+0.045}$ & $1.198_{-0.062}^{+0.065}$ \\
         NGC0507 &    1 & $1.078_{-0.024}^{+0.029}$ & $0.374_{-0.069}^{+0.087}$ & $1.077_{-0.028}^{+0.040}$ & $0.353_{-0.079}^{+0.115}$ \\
         	 &    2 & $1.120_{-0.009}^{+0.009}$ & $0.567_{-0.043}^{+0.048}$ & $1.477_{-0.023}^{+0.021}$ & $0.934_{-0.091}^{+0.102}$ \\
            A262 &    1 & $0.931_{-0.021}^{+0.019}$ & $0.311_{-0.056}^{+0.072}$ & $1.432_{-0.015}^{+0.014}$ & $1.012_{-0.086}^{+0.095}$ \\
            	 &    2 & $1.342_{-0.007}^{+0.006}$ & $0.741_{-0.030}^{+0.032}$ & $1.672_{-0.017}^{+0.016}$ & $0.824_{-0.043}^{+0.045}$ \\
            	 &    3 & $\cdots$		    & $\cdots$			& $1.189_{-0.012}^{+0.011}$ & $0.476_{-0.038}^{+0.042}$ \\
           A3112 &    1 & $3.055_{-0.037}^{+0.041}$ & $1.447_{-0.067}^{+0.069}$ & $3.672_{-0.051}^{+0.053}$ & $1.243_{-0.064}^{+0.067}$ \\
  Fornax cluster &    1 & $0.940_{-0.003}^{+0.003}$ & $0.719_{-0.028}^{+0.029}$ & $0.988_{-0.003}^{+0.003}$ & $0.832_{-0.033}^{+0.035}$ \\
  	 	 &    2 & $1.008_{-0.012}^{+0.012}$ & $0.434_{-0.047}^{+0.053}$ & $1.005_{-0.010}^{+0.010}$ & $0.823_{-0.095}^{+0.114}$ \\
      2A0335+096 &    1 & $2.127_{-0.009}^{+0.009}$ & $0.716_{-0.014}^{+0.014}$ & $2.212_{-0.013}^{+0.013}$ & $0.757_{-0.020}^{+0.020}$ \\
            A478 &    1 & $4.560_{-0.085}^{+0.095}$ & $0.648_{-0.048}^{+0.050}$ & $5.200_{-0.112}^{+0.111}$ & $0.599_{-0.050}^{+0.052}$ \\
  RXJ0419.6+0225 &    1 & $1.133_{-0.004}^{+0.004}$ & $0.541_{-0.021}^{+0.022}$ & $1.266_{-0.007}^{+0.007}$ & $0.593_{-0.029}^{+0.031}$ \\
     EXO0422-086 &    1 & $2.560_{-0.036}^{+0.038}$ & $0.920_{-0.061}^{+0.065}$ & $2.636_{-0.076}^{+0.078}$ & $1.252_{-0.159}^{+0.180}$ \\
     		 &    2 & $\cdots$		    & $\cdots$			& $3.225_{-0.068}^{+0.073}$ & $0.887_{-0.087}^{+0.093}$ \\
            A496 &    1 & $2.098_{-0.024}^{+0.024}$ & $1.220_{-0.056}^{+0.059}$ & $3.381_{-0.032}^{+0.032}$ & $1.216_{-0.039}^{+0.041}$ \\
            	 &    2 & $3.338_{-0.071}^{+0.071}$ & $1.112_{-0.077}^{+0.082}$ & $\cdots$		    & $\cdots$			\\
	Hydra A		 & 1	& $2.832_{-0.020}^{+0.020}$ & $0.726_{-0.020}^{+0.020}$	& $3.416_{-0.029}^{+0.030}$	& $0.675_{-0.023}^{+0.022}$	\\
				 & 2	& $3.235_{-0.073}^{+0.080}$	& $0.685_{-0.064}^{+0.067}$	& $\cdots$		    & $\cdots$			\\
            MKW4 &    1 & $1.590_{-0.029}^{+0.027}$ & $2.377_{-0.328}^{+0.400}$ & $1.607_{-0.026}^{+0.033}$ & $2.643_{-0.372}^{+0.456}$ \\
            	 &    2 & $1.813_{-0.040}^{+0.051}$ & $2.018_{-0.248}^{+0.300}$ & $2.007_{-0.051}^{+0.059}$ & $1.458_{-0.143}^{+0.159}$ \\
         NGC4636 &    1 & $0.642_{-0.005}^{+0.003}$ & $0.376_{-0.014}^{+0.015}$ & $0.655_{-0.005}^{+0.005}$ & $0.705_{-0.048}^{+0.053}$ \\
           A3526 &    1 & $0.977_{-0.002}^{+0.002}$ & $0.321_{-0.006}^{+0.007}$ & $1.125_{-0.003}^{+0.003}$ & $0.303_{-0.007}^{+0.007}$ \\
           	 &    2 & $1.414_{-0.001}^{+0.001}$ & $0.941_{-0.009}^{+0.009}$ & $1.792_{-0.003}^{+0.003}$ & $1.526_{-0.015}^{+0.015}$ \\
           A1644 &    1 & $2.653_{-0.053}^{+0.052}$ & $1.342_{-0.105}^{+0.114}$ & $4.471_{-0.183}^{+0.197}$ & $0.648_{-0.120}^{+0.131}$ \\
         NGC5044 &    1 & $0.845_{-0.002}^{+0.002}$ & $0.523_{-0.013}^{+0.013}$ & $0.902_{-0.003}^{+0.003}$ & $0.479_{-0.014}^{+0.015}$ \\
         	 &    2 & $0.885_{-0.002}^{+0.002}$ & $0.602_{-0.014}^{+0.015}$ & $1.018_{-0.002}^{+0.001}$ & $0.553_{-0.011}^{+0.009}$ \\
         	 &    3 & $0.914_{-0.008}^{+0.009}$ & $0.715_{-0.073}^{+0.089}$ & $0.922_{-0.004}^{+0.004}$ & $0.605_{-0.030}^{+0.033}$ \\
           A1795 &    1 & $3.601_{-0.014}^{+0.013}$ & $0.768_{-0.013}^{+0.013}$ & $4.921_{-0.022}^{+0.022}$ & $0.683_{-0.013}^{+0.013}$ \\
           A3581 &    1 & $1.070_{-0.011}^{+0.011}$ & $0.243_{-0.020}^{+0.022}$ & $1.400_{-0.030}^{+0.028}$ & $0.328_{-0.044}^{+0.049}$ \\
           	 &    2 & $1.399_{-0.009}^{+0.008}$ & $0.693_{-0.029}^{+0.030}$ & $1.684_{-0.016}^{+0.016}$ & $0.910_{-0.047}^{+0.050}$ \\
 RXCJ1504.1-0248 &    1 & $4.597_{-0.059}^{+0.061}$ & $0.490_{-0.025}^{+0.025}$ & $5.146_{-0.075}^{+0.082}$ & $0.389_{-0.027}^{+0.028}$ \\
           A2029 &    1 & $5.667_{-0.112}^{+0.119}$ & $1.353_{-0.096}^{+0.101}$ & $7.395_{-0.102}^{+0.103}$ & $0.819_{-0.043}^{+0.044}$ \\
           	 &    2 & $6.952_{-0.084}^{+0.084}$ & $0.898_{-0.040}^{+0.041}$ & $\cdots$		    & $\cdots$			\\
           A2052 &    1 & $2.009_{-0.006}^{+0.006}$ & $0.703_{-0.009}^{+0.009}$ & $1.662_{-0.014}^{+0.014}$ & $0.627_{-0.028}^{+0.029}$ \\
           	 &    2 & $1.800_{-0.015}^{+0.015}$ & $1.026_{-0.047}^{+0.049}$ & $3.051_{-0.017}^{+0.017}$ & $0.896_{-0.017}^{+0.015}$ \\
           	 &    3 & $\cdots$		    & $\cdots$			& $2.178_{-0.040}^{+0.034}$ & $1.198_{-0.069}^{+0.068}$ \\
           MKW3S &    1 & $3.086_{-0.044}^{+0.045}$ & $1.197_{-0.063}^{+0.066}$ & $3.872_{-0.068}^{+0.074}$ & $1.018_{-0.073}^{+0.076}$ \\
           A2199 &    1 & $2.478_{-0.026}^{+0.026}$ & $1.347_{-0.054}^{+0.057}$ & $3.271_{-0.043}^{+0.043}$ & $1.160_{-0.053}^{+0.055}$ \\
           	 &    2 & $3.457_{-0.052}^{+0.052}$ & $1.050_{-0.056}^{+0.059}$ & $3.791_{-0.049}^{+0.048}$ & $0.927_{-0.047}^{+0.048}$ \\
           A2204 &    1 & $4.871_{-0.065}^{+0.069}$ & $1.031_{-0.054}^{+0.056}$ & $5.427_{-0.088}^{+0.098}$ & $0.700_{-0.050}^{+0.052}$ \\
\enddata
%\tablenotetext{a}{Identification number of distinct perturbed (positive and negative excess) regions identified in the \Chandra ~X-ray residual image.}
%\tablenotetext{b}{Observed ICM quantities correspond to the projected ones.}
\end{deluxetable*}

%% file: table/table_kTb.tex
\begin{deluxetable*}{lccccc}
\addtocounter{table}{-1}
\tablecaption{
Continued.
}
\tablehead{
\colhead{Cluster} &
\colhead{ID\tablenotemark{a}}	&
\multicolumn{2}{c}{Positive excess region\tablenotemark{b}}	& 
\multicolumn{2}{c}{Negative excess region\tablenotemark{b}}	\\
\colhead{}		&
\colhead{}		&
\colhead{$kT$	(keV)}			&
\colhead{Abundance (\ZO)}	&
\colhead{$kT$ (keV)}	&
\colhead{Abundance (\ZO)}
}
\startdata
          AS1101 &    1 & $2.530_{-0.028}^{+0.028}$ & $0.623_{-0.033}^{+0.035}$ & $2.500_{-0.025}^{+0.025}$ & $0.600_{-0.028}^{+0.029}$ \\
           A2597 &    1 & $2.703_{-0.022}^{+0.020}$ & $0.578_{-0.022}^{+0.022}$ & $2.914_{-0.035}^{+0.035}$ & $0.588_{-0.032}^{+0.034}$ \\
           	 &    2 & $3.211_{-0.022}^{+0.022}$ & $0.589_{-0.020}^{+0.020}$ & $2.884_{-0.021}^{+0.023}$ & $0.543_{-0.019}^{+0.018}$ \\
           	 &    3 & $2.898_{-0.082}^{+0.087}$ & $0.627_{-0.082}^{+0.088}$ & $3.817_{-0.056}^{+0.057}$ & $0.512_{-0.038}^{+0.040}$ \\
           A4059 &    1 & $1.570_{-0.028}^{+0.027}$ & $0.779_{-0.075}^{+0.083}$ & $3.374_{-0.053}^{+0.053}$ & $1.308_{-0.068}^{+0.071}$ \\
           	 &    2 & $2.481_{-0.042}^{+0.041}$ & $1.783_{-0.101}^{+0.108}$ & $4.602_{-0.125}^{+0.132}$ & $0.808_{-0.084}^{+0.090}$ \\
\enddata
\tablenotetext{a}{Identification number of distinct perturbed (positive and negative excess) regions identified in the \Chandra ~X-ray residual image.}
\tablenotetext{b}{Observed projected ICM quantities.}
\end{deluxetable*}

%% file: table/table_nPKa.tex
\begin{deluxetable*}{lccccccc}
\tablecaption{
Electron number density ($n_{\rm e}$), pressure ($p_{\rm e}$), and entropy ($K_{\rm e}$) measured in each of the perturbed regions identified in X-ray residual images. \label{tab:npe}
}
\tablehead{
\colhead{Cluster}	&
\colhead{ID\tablenotemark{a}}	&
\multicolumn{3}{c}{Positive excess region\tablenotemark{b}}	&
\multicolumn{3}{c}{Negative excess region\tablenotemark{b}}	\\
\colhead{}	&
\colhead{}	&
\colhead{$n_{\rm e}$\tablenotemark{c}}	&
\colhead{$p_{\rm e}$\tablenotemark{d}}	&
\colhead{$K_{\rm e}$\tablenotemark{e}}	& 
\colhead{$n_{\rm e}$\tablenotemark{c}}	& 
\colhead{$p_{\rm e}$\tablenotemark{d}}	&
\colhead{$K_{\rm e}$\tablenotemark{e}}
}
\startdata
             A85 & 1 & $1.076_{-0.031}^{+0.031}$ & $2.712_{-0.109}^{+0.116}$ & $51.744_{-1.762}^{+1.920}$ & $0.662_{-0.024}^{+0.025}$ & $2.457_{-0.159}^{+0.167}$ &$105.379_{-6.147}^{+6.567}$ \\
             	 & 2 & $0.732_{-0.006}^{+0.006}$ & $2.861_{-0.049}^{+0.049}$ &$103.634_{-1.651}^{+1.649}$ & $0.680_{-0.007}^{+0.007}$ & $2.782_{-0.060}^{+0.065}$ &$114.041_{-2.301}^{+2.505}$ \\
            A133 & 1 & $0.525_{-0.005}^{+0.005}$ & $1.189_{-0.015}^{+0.015}$ & $74.932_{-0.823}^{+0.819}$ & $0.369_{-0.003}^{+0.003}$ & $1.183_{-0.019}^{+0.019}$ &$134.043_{-1.983}^{+2.033}$ \\
         NGC0507 & 1 & $0.257_{-0.015}^{+0.016}$ & $0.278_{-0.018}^{+0.019}$ & $57.415_{-2.632}^{+2.876}$ & $0.255_{-0.018}^{+0.020}$ & $0.274_{-0.021}^{+0.024}$ & $57.719_{-3.164}^{+3.668}$ \\
         	 & 2 & $0.174_{-0.005}^{+0.005}$ & $0.194_{-0.005}^{+0.005}$ & $77.558_{-1.513}^{+1.521}$ & $0.103_{-0.003}^{+0.003}$ & $0.152_{-0.005}^{+0.005}$ &$144.635_{-3.597}^{+3.554}$ \\
            A262 & 1 & $0.646_{-0.042}^{+0.044}$ & $0.602_{-0.042}^{+0.042}$ & $26.823_{-1.317}^{+1.327}$ & $0.342_{-0.009}^{+0.010}$ & $0.489_{-0.014}^{+0.015}$ & $63.135_{-1.335}^{+1.336}$ \\
            	 & 2 & $0.412_{-0.005}^{+0.005}$ & $0.553_{-0.007}^{+0.007}$ & $52.228_{-0.504}^{+0.503}$ & $0.269_{-0.003}^{+0.003}$ & $0.451_{-0.007}^{+0.007}$ & $86.375_{-1.116}^{+1.101}$ \\
            	 & 3 & $\cdots$ 		 & $\cdots$		     & $\cdots$			  & $0.460_{-0.011}^{+0.011}$ & $0.546_{-0.014}^{+0.014}$ & $43.001_{-0.815}^{+0.814}$ \\
           A3112 & 1 & $0.970_{-0.008}^{+0.009}$ & $2.962_{-0.044}^{+0.047}$ & $67.187_{-0.908}^{+0.981}$ & $0.829_{-0.007}^{+0.006}$ & $3.046_{-0.049}^{+0.050}$ & $89.621_{-1.334}^{+1.385}$ \\
  Fornax cluster & 1 & $0.384_{-0.006}^{+0.006}$ & $0.361_{-0.006}^{+0.006}$ & $38.340_{-0.419}^{+0.422}$ & $0.332_{-0.005}^{+0.005}$ & $0.328_{-0.005}^{+0.005}$ & $44.377_{-0.499}^{+0.502}$ \\
  	 	 & 2 & $0.255_{-0.010}^{+0.010}$ & $0.257_{-0.010}^{+0.010}$ & $54.033_{-1.518}^{+1.534}$ & $0.372_{-0.018}^{+0.019}$ & $0.374_{-0.019}^{+0.019}$ & $41.884_{-1.429}^{+1.456}$ \\
      2A0335+096 & 1 & $0.779_{-0.003}^{+0.003}$ & $1.657_{-0.009}^{+0.009}$ & $54.129_{-0.261}^{+0.261}$ & $0.618_{-0.003}^{+0.003}$ & $1.368_{-0.010}^{+0.011}$ & $65.686_{-0.444}^{+0.443}$ \\
            A478 & 1 & $1.174_{-0.008}^{+0.008}$ & $5.353_{-0.106}^{+0.118}$ & $88.287_{-1.693}^{+1.885}$ & $1.021_{-0.007}^{+0.007}$ & $5.309_{-0.120}^{+0.119}$ &$110.481_{-2.434}^{+2.406}$ \\
  RXJ0419.6+0225 & 1 & $0.379_{-0.005}^{+0.005}$ & $0.430_{-0.006}^{+0.006}$ & $46.564_{-0.440}^{+0.441}$ & $0.294_{-0.004}^{+0.004}$ & $0.372_{-0.006}^{+0.006}$ & $61.743_{-0.710}^{+0.710}$ \\
     EXO0422-086 & 1 & $0.487_{-0.006}^{+0.006}$ & $1.247_{-0.023}^{+0.023}$ & $89.138_{-1.416}^{+1.483}$ & $0.705_{-0.018}^{+0.018}$ & $1.859_{-0.072}^{+0.073}$ & $71.683_{-2.409}^{+2.456}$ \\
     		 & 2 & $\cdots$			 & $\cdots$		     & $\cdots$			  & $0.414_{-0.006}^{+0.006}$ & $1.336_{-0.034}^{+0.035}$ &$125.023_{-2.872}^{+3.067}$ \\
            A496 & 1 & $0.801_{-0.008}^{+0.008}$ & $1.681_{-0.026}^{+0.026}$ & $52.408_{-0.699}^{+0.693}$ & $0.450_{-0.002}^{+0.002}$ & $1.520_{-0.016}^{+0.016}$ &$124.116_{-1.247}^{+1.245}$ \\
            	 & 2 & $0.454_{-0.005}^{+0.005}$ & $1.516_{-0.036}^{+0.036}$ &$121.677_{-2.706}^{+2.700}$ & $\cdots$		      & $\cdots$		  &$\cdots$		       \\
		Hydra A & 1 & $0.915_{-0.003}^{+0.003}$	& $2.591_{-0.020}^{+0.020}$ & $64.722_{-0.480}^{+0.481}$ & $0.820_{-0.003}^{+0.003}$ & $2.801_{-0.026}^{+0.026}$ & $84.005_{-0.739}^{+0.761}$ \\
				 & 2 & $0.941_{-0.010}^{+0.010}$ & $3.045_{-0.076}^{+0.081}$ & $72.548_{-1.718}^{+1.860}$	& $\cdots$	& $\cdots$	& $\cdots$	\\
            MKW4 & 1 & $0.294_{-0.016}^{+0.016}$ & $0.468_{-0.026}^{+0.027}$ & $77.456_{-3.100}^{+3.104}$ & $0.261_{-0.015}^{+0.015}$ & $0.419_{-0.025}^{+0.026}$ & $84.852_{-3.464}^{+3.695}$ \\
            	 & 2 & $0.167_{-0.006}^{+0.007}$ & $0.303_{-0.014}^{+0.015}$ &$128.744_{-4.387}^{+4.974}$ & $0.142_{-0.003}^{+0.003}$ & $0.286_{-0.010}^{+0.011}$ &$158.644_{-4.770}^{+5.304}$ \\
         NGC4636 & 1 & $0.524_{-0.007}^{+0.008}$ & $0.336_{-0.005}^{+0.005}$ & $21.291_{-0.261}^{+0.242}$ & $0.287_{-0.009}^{+0.009}$ & $0.188_{-0.006}^{+0.006}$ & $32.412_{-0.705}^{+0.715}$ \\
           A3526 & 1 & $1.210_{-0.007}^{+0.007}$ & $1.182_{-0.008}^{+0.008}$ & $18.535_{-0.087}^{+0.087}$ & $0.947_{-0.006}^{+0.006}$ & $1.066_{-0.007}^{+0.007}$ & $25.135_{-0.118}^{+0.118}$ \\
           	 & 2 & $0.591_{-0.002}^{+0.002}$ & $0.836_{-0.003}^{+0.003}$ & $43.266_{-0.096}^{+0.096}$ & $0.433_{-0.001}^{+0.001}$ & $0.775_{-0.003}^{+0.003}$ & $67.494_{-0.167}^{+0.167}$ \\
           A1644 & 1 & $0.272_{-0.004}^{+0.004}$ & $0.721_{-0.019}^{+0.019}$ &$136.234_{-3.090}^{+3.073}$ & $0.165_{-0.003}^{+0.003}$ & $0.737_{-0.032}^{+0.034}$ &$320.225_{-3.512}^{+14.474}$ \\
         NGC5044 & 1 & $0.378_{-0.004}^{+0.004}$ & $0.320_{-0.003}^{+0.003}$ & $34.806_{-0.242}^{+0.243}$ & $0.336_{-0.004}^{+0.004}$ & $0.303_{-0.004}^{+0.003}$ & $40.206_{-0.322}^{+0.318}$ \\
         	 & 2 & $0.246_{-0.002}^{+0.002}$ & $0.218_{-0.002}^{+0.002}$ & $48.580_{-0.320}^{+0.321}$ & $0.200_{-0.001}^{+0.001}$ & $0.204_{-0.001}^{+0.001}$ & $64.105_{-0.292}^{+0.298}$ \\
         	 & 3 & $0.353_{-0.016}^{+0.016}$ & $0.322_{-0.015}^{+0.015}$ & $39.474_{-1.231}^{+1.229}$ & $0.267_{-0.005}^{+0.005}$ & $0.246_{-0.005}^{+0.005}$ & $47.906_{-0.677}^{+0.682}$ \\
           A1795 & 1 & $0.956_{-0.002}^{+0.002}$ & $3.442_{-0.015}^{+0.014}$ & $79.963_{-0.321}^{+0.314}$ & $0.706_{-0.001}^{+0.001}$ & $3.473_{-0.016}^{+0.016}$ &$133.746_{-0.616}^{+0.617}$ \\
           A3581 & 1 & $0.699_{-0.015}^{+0.015}$ & $0.748_{-0.018}^{+0.018}$ & $29.256_{-0.527}^{+0.527}$ & $0.577_{-0.016}^{+0.016}$ & $0.808_{-0.028}^{+0.028}$ & $43.530_{-1.223}^{+1.197}$ \\
           	 & 2 & $0.393_{-0.005}^{+0.005}$ & $0.549_{-0.007}^{+0.007}$ & $56.223_{-0.550}^{+0.548}$ & $0.322_{-0.004}^{+0.004}$ & $0.543_{-0.009}^{+0.009}$ & $77.193_{-0.982}^{+0.969}$ \\
 RXCJ1504.1-0248 & 1 & $2.577_{-0.011}^{+0.011}$ &$11.847_{-0.159}^{+0.164}$ & $52.680_{-0.688}^{+0.710}$ & $2.124_{-0.009}^{+0.009}$ &$10.932_{-0.166}^{+0.181}$ & $67.088_{-0.995}^{+1.087}$ \\
           A2029 & 1 & $1.265_{-0.010}^{+0.010}$ & $7.168_{-0.153}^{+0.161}$ &$104.393_{-2.136}^{+2.259}$ & $0.837_{-0.003}^{+0.003}$ & $6.189_{-0.088}^{+0.088}$ &$179.379_{-2.519}^{+2.520}$ \\
           	 & 2 & $0.818_{-0.003}^{+0.003}$ & $5.686_{-0.071}^{+0.071}$ &$171.230_{-2.094}^{+2.094}$ & $\cdots$ 		      & $\cdots$		  & $\cdots$		       \\
           A2052 & 1 & $0.624_{-0.001}^{+0.001}$ & $1.253_{-0.004}^{+0.004}$ & $59.265_{-0.186}^{+0.186}$ & $0.618_{-0.006}^{+0.006}$ & $1.027_{-0.013}^{+0.013}$ & $49.359_{-0.519}^{+0.515}$ \\
           	 & 2 & $0.688_{-0.008}^{+0.008}$ & $1.238_{-0.017}^{+0.017}$ & $49.738_{-0.562}^{+0.557}$ & $0.366_{-0.001}^{+0.001}$ & $1.118_{-0.007}^{+0.007}$ &$128.381_{-0.747}^{+0.756}$ \\
           	 & 3 & $\cdots$			 & $\cdots$		     & $\cdots$			  & $0.509_{-0.006}^{+0.005}$ & $1.109_{-0.024}^{+0.021}$ & $73.565_{-1.470}^{+1.271}$ \\
           MKW3S & 1 & $0.447_{-0.004}^{+0.004}$ & $1.381_{-0.023}^{+0.024}$ &$113.651_{-1.761}^{+1.806}$ & $0.354_{-0.003}^{+0.003}$ & $1.369_{-0.027}^{+0.029}$ &$166.826_{-3.111}^{+3.350}$ \\
\enddata
%\tablenotetext{a}{Identification number of distinct perturbed (positive and negative excess) regions identified in the \Chandra ~X-ray residual image.}
%\tablenotetext{b}{Observed ICM quantities correspond to the projected one.}
%\tablenotetext{c}{Electron number density in units of $10^{-2}$\,cm\,$^{-3}$\,($L$/1\,Mpc)$^{-1/2}$.}
%\tablenotetext{d}{Electron pressure in units of $10^{-2}$\,keV\,cm$^{-3}$\,($L$/1\,Mpc)$^{-1/2}$.}
%\tablenotetext{e}{Electron entropy in units of keV\,cm$^2$\,($L$/1\,Mpc)$^{1/3}$.}
\end{deluxetable*}

%% file: table/table_nPKb.tex
\begin{deluxetable*}{lccccccc}
\addtocounter{table}{-1}
\tablecaption{
Continued.
}
\tablehead{
\colhead{Cluster}	&
\colhead{ID\tablenotemark{a}}	&
\multicolumn{3}{c}{Positive excess region\tablenotemark{b}}	&
\multicolumn{3}{c}{Negative excess region\tablenotemark{b}}	\\
\colhead{}	&
\colhead{}	&
\colhead{$n_{\rm e}$\tablenotemark{c}}	&
\colhead{$p_{\rm e}$\tablenotemark{d}}	&
\colhead{$K_{\rm e}$\tablenotemark{e}}	& 
\colhead{$n_{\rm e}$\tablenotemark{c}}	& 
\colhead{$p_{\rm e}$\tablenotemark{d}}	&
\colhead{$K_{\rm e}$\tablenotemark{e}}
}
\startdata
           A2199 & 1 & $0.791_{-0.007}^{+0.007}$ & $1.960_{-0.027}^{+0.027}$ & $62.408_{-0.753}^{+0.753}$ & $0.601_{-0.004}^{+0.004}$ & $1.966_{-0.029}^{+0.029}$ & $98.946_{-1.379}^{+1.370}$ \\
           	 & 2 & $0.593_{-0.004}^{+0.004}$ & $2.052_{-0.034}^{+0.034}$ &$105.473_{-1.663}^{+1.671}$ & $0.498_{-0.003}^{+0.003}$ & $1.890_{-0.027}^{+0.027}$ &$129.906_{-1.742}^{+1.736}$ \\
           A2204 & 1 & $1.709_{-0.010}^{+0.010}$ & $8.323_{-0.122}^{+0.127}$ & $73.419_{-1.025}^{+1.076}$ & $1.366_{-0.007}^{+0.007}$ & $7.414_{-0.126}^{+0.140}$ & $94.956_{-1.569}^{+1.747}$ \\
          AS1101 & 1 & $0.678_{-0.005}^{+0.005}$ & $1.714_{-0.023}^{+0.023}$ & $70.659_{-0.866}^{+0.868}$ & $0.499_{-0.003}^{+0.003}$ & $1.248_{-0.015}^{+0.015}$ & $85.576_{-0.913}^{+0.915}$ \\
           A2597 & 1 & $1.244_{-0.006}^{+0.006}$ & $3.362_{-0.031}^{+0.030}$ & $50.361_{-0.431}^{+0.409}$ & $1.057_{-0.007}^{+0.007}$ & $3.080_{-0.042}^{+0.042}$ & $60.506_{-0.774}^{+0.773}$ \\
           	 & 2 & $0.781_{-0.003}^{+0.003}$ & $2.507_{-0.019}^{+0.019}$ & $81.601_{-0.596}^{+0.580}$ & $0.923_{-0.003}^{+0.004}$ & $2.663_{-0.022}^{+0.023}$ & $65.518_{-0.513}^{+0.547}$ \\
           	 & 3 & $0.948_{-0.015}^{+0.015}$ & $2.749_{-0.089}^{+0.094}$ & $64.691_{-1.948}^{+2.065}$ & $0.592_{-0.004}^{+0.004}$ & $2.261_{-0.036}^{+0.036}$ &$116.599_{-1.783}^{+1.808}$ \\
           A4059 & 1 & $0.612_{-0.015}^{+0.015}$ & $0.961_{-0.029}^{+0.029}$ & $46.918_{-1.129}^{+1.106}$ & $0.334_{-0.003}^{+0.003}$ & $1.128_{-0.020}^{+0.020}$ &$150.907_{-2.528}^{+2.523}$ \\
           	 & 2 & $0.432_{-0.006}^{+0.006}$ & $1.071_{-0.023}^{+0.023}$ & $93.577_{-1.784}^{+1.768}$ & $0.260_{-0.002}^{+0.002}$ & $1.198_{-0.035}^{+0.036}$ &$243.140_{-6.794}^{+7.114}$ \\
\enddata
\tablenotetext{a}{Identification number of distinct perturbed (positive and negative excess) regions detected in the \Chandra\ X-ray residual image.}
\tablenotetext{b}{Observed projected ICM quantities.}
\tablenotetext{c}{Electron number density in units of $10^{-2}$\,cm\,$^{-3}$\,($L$/1\,Mpc)$^{-1/2}$.}
\tablenotetext{d}{Electron pressure in units of $10^{-2}$\,keV\,cm$^{-3}$\,($L$/1\,Mpc)$^{-1/2}$.}
\tablenotetext{e}{Electron entropy in units of keV\,cm$^2$\,($L$/1\,Mpc)$^{1/3}$.}
\end{deluxetable*}